%% file: ms_arxiv.tex
\documentclass[fleqn,usenatbib]{mnras}

\usepackage[T1]{fontenc}
\usepackage{ae,aecompl}

\usepackage{graphicx}	
\usepackage{amsmath}	
\usepackage{amssymb}	
\usepackage{threeparttable}
\usepackage{multirow}
\usepackage{url}
\usepackage{multicol}

\usepackage{pdflscape}    
\usepackage[dvipsnames]{xcolor} 

\usepackage{newtxtext,newtxmath}

\newcommand{\mum}{\ifmmode{\rm \mu m}\else{$\mu$m}\fi}
\newcommand{\Msun}{\ensuremath{{\rm M}_{\odot}}}

\newcommand{\kms}{km~s\ensuremath{^{-1}}}                          

\newcommand{\Msunyr}{\ensuremath{{\rm M}_{\odot} \, {\rm yr}^{-1}}}

\newcommand{\Lsol}{L$_{\odot}$}

\def\eg{{e.g.,~}}

\DeclareUnicodeCharacter{2212}{\textendash}

\newcommand{\Hii}{H\,{\sc ii}}  
\newcommand{\brg}{Br~$\gamma$} 

\newcommand{\pp}{$\phantom{1}$}

\title[YSOs in NGC 346]{Near-infrared spectroscopy of embedded protostars in the massive metal-poor star forming region NGC 346}

\author[O. C. Jones et al.]{O.~C.~Jones$^{1}$\thanks{E-mail: olivia.jones@stfc.ac.uk},
M.~Reiter$^{1,2}$,
R.~Sanchez-Janssen$^{1}$,
C.~J.~Evans$^{1,3}$,
C.~S.~Robertson$^{4}$,
\newauthor
M.~Meixner$^{5}$
and
B.~Ochsendorf$^{6}$
\\
$^{1}$UK Astronomy Technology Centre, Royal Observatory, Blackford Hill, Edinburgh, EH9 3HJ, UK\\\(\)
$^{2}$European Southern Observatory, Karl-Schwarzschild-Strasse 2, D-85748 Garching bei M\"{u}nchen, Germany\\
$^{3}$European Space Agency (ESA), ESA Office, Space Telescope Science Institute, 3700 San Martin Drive, Baltimore, MD 21218, USA\\
$^{4}$Institute for Astronomy, University of Edinburgh, Blackford Hill, Edinburgh EH9 3HJ, UK \\\(\)
$^{5}$SOFIA-USRA, NASA Ames Research Center, MS 232-12, Moffett Field, CA 94035, USA \\
$^{6}$ Department of Physics and Astronomy, Johns Hopkins University, 3400 North Charles Street, Baltimore, MD 21218, USA
}

\date{Accepted XXX. Received YYY; in original form \today}

\pubyear{2022}

\begin{document}
\label{firstpage}
\pagerange{\pageref{firstpage}--\pageref{lastpage}}
\maketitle

\begin{abstract}
We present medium-resolution (R $\sim$ 4000) YJ, H \& K band spectroscopy of candidate young stellar objects (YSOs) in NGC~346, the most active star-formation region in the metal-poor (Z = 1/5 Z$_{\sun}$) Small Magellanic Cloud. 
The spectra were obtained with the KMOS (K-Band Multi Object Spectrograph) integral field instrument on the Very Large Telescope.
 From our initial sample of 18 candidate high-mass YSOs previously identified from mid-IR photometry and radiative transfer model fits to their spectral energy distributions, approximately half were resolved into multiple components by our integral-field data. In total we detect 30 continuum sources and extract reliable spectra for 12 of these objects.
The spectra show various features including hydrogen recombination lines, and lines from H$_2$, He~{\sc i} and [Fe~{\sc ii}], which are indicative of accretion, discs and outflowing material in massive YSOs.  We spectroscopically confirm the youthful nature of nine YSO candidates, and identify two others as OB stars.    
All of the confirmed YSOs have Br$\gamma$ in emission, but no emission is seen from the CO bandhead, despite other disc tracers present in the spectra.  He\,{\sc i}~1.083 $\mu$m emission is also detected at appreciably higher rates than for the Galaxy.
\end{abstract}




\section{Introduction}
\label{sec:intro}

The Small Magellanic Cloud (SMC), at a distance of $\sim$62\,kpc \citep{deGrijs2015} and with a metallicity of $\sim$1/5 solar \citep{Peimbert2000}, is actively forming stars at a rate of $\sim$0.05 ${\rm M}_{\odot} \, {\rm yr}^{-1}$ \citep{Wilke2004}. Its interstellar medium (ISM) differs significantly from that of the Milky Way, with a lower dust abundance and extreme UV extinction curve variations (which has a steep UV rise and
no 2175 \AA~bump in the Bar; \citealt{Gordon2003}), providing a local proxy for a starburst galaxy reminiscent of the early universe 
\citep{Dimaratos2015}. 
Its most active (and brightest) star-forming region, NGC~346 (also known as N66, or DEM~103; \citealt{Henize1956}), is located at the northern end of the SMC bar.  Its H$\alpha$ luminosity is 60 times brighter than Orion \citep{Kennicutt1984} and over the last 100 Myr it is forming stars at a rate of $1.4\times10^{-8}$ \Msunyr pc$^{-2}$ \citep{Cignoni2011}.
This giant H\,{\sc ii} region is young ($\sim$3 Myr; \citealt{Bouret2003}), containing over 30 O-type stars (more than half of those known in the entire SMC; \citealt{Massey1989, Evans2006}) with masses in the range of 35--100 M$_\odot$  \citep[e.g.,][]{Dufton2019}. {\em Hubble Space Telescope (HST)} imaging revealed thousands of candidate low-mass (0.6--3 M$_\odot$), pre-main-sequence (PMS) stars distributed amidst the young stars of NGC~346, and scores of small compact clusters connected by gas and dust filaments \citep{Nota2006, Sabbi2007, Hennekemper2008, DeMarchi2011}. Combined, this high-resolution, multi-wavelength imaging indicates that NGC~346 has experienced an intriguing series of (presumed) sequential star-formation episodes, with the region shaped by feedback from its most massive stars.  

NGC~346 has been studied extensively with {\em Spitzer} and {\em Herschel}\ \citep{Simon2007, Sewilo2013, Seale2014, Ruffle2015}, to identify metal-poor, young stellar objects (YSOs) in the very early stage of formation, resulting in the discovery of approximately 100 stage I, II, and III YSO candidates. 
Forming within the past $\sim$1 Myr, these infrared-bright, candidate embedded YSOs in NGC~346 are thought to have stellar masses ranging from  1.5--17 M$_\odot$ \citep{Simon2007}; most are associated with H$\alpha$ emission with diverse morphologies and many are found at the tip of, or inside, dusty pillars \citep{Sewilo2013}. These observations suggest star formation is ongoing throughout the complex at a rate of $>3.2 \times 10^{-3}$ ${\rm M}_{\odot} \, {\rm yr}^{-1}$ \citep{Simon2007}. 

Spectroscopic confirmation of massive YSOs in the SMC is currently limited to a few tens of sources \citep{Oliveira2011, Oliveira2013, Oliveira2019, Ruffle2015, Ward2017, Rubio2018, Reiter2019}. Of these, seven are located in NGC~346 \citep{Ruffle2015, Rubio2018}, with six confirmed as YSOs based on the properties of their {\em Spitzer} InfraRed Spectrograph (IRS) spectra. Near-IR spectroscopy is currently only available for three massive YSOs in NGC~346 \citep{Rubio2018}.  

Near-IR integral-field spectroscopy of candidate YSOs in the SMC revealed a wide range of morphological and spectral properties, with seven sources exhibiting extended H$_2$ emission indicative of outflows in early-stage YSOs \citep{Ward2017}. In this wavelength regime, massive YSOs can be conclusively identified due to the wealth of atomic and molecular emission lines in their near-IR spectra. Ionized atomic emission lines measure the excitation of the YSO and surrounding ISM, while key features trace accretion in the inner disc region (e.g.,~Br$\gamma$), the presence of outflows (e.g.,~H$_{2}$ and [Fe{\sc ii}]) and circumstellar discs (via the CO bandhead emission and fluorescent Fe~{\sc ii}).  

Medium resolution, near-IR spectroscopy of candidate YSOs in the Clouds has been conducted by \citet{Ward2016, Ward2017}, \citet{Rubio2018}, \citet{Reiter2019}, \citet{vanGelder2020} and \citet{Sewilo2022}. In these samples the Br$\gamma$ line has been detected at a high rate, but no conclusive relation has been found between Br$\gamma$ luminosity and metallicity, although \citet{Ward2017} suggested their YSO sample shows tentative evidence of increased accretion rates in the SMC compared to the Milky Way.

In this paper we examine the fundamental properties of massive star formation at sub-solar metallicity in NGC~346 and explore variations within the Magellanic Clouds \citep[c.f.][]{Ward2016} and compared to the Milky Way \citep{Cooper2013}. We present near-IR integral-field spectroscopy of IR-bright massive YSO candidates in NGC~346 observed with the K-band Multi-Object Spectrograph (KMOS; \citealt{Sharples2013}) on the Very Large Telescope (VLT) at the European Southern Observatory (ESO). The aim is to confirm their massive YSO nature using near-IR emission features. 
The sample selection, KMOS observations and data reduction are described in Section~\ref{sec:observations}. 
Section~\ref{sec:results} presents the main results and our analysis, including an overview of the spectra and their emission line fluxes. 
In Section~\ref{sec:discussion} we discuss the properties of massive YSOs in NGC~346, and compare our results to other near-IR spectroscopic samples of YSOs in the Galaxy and Magellanic Clouds. Finally, we present a summary of our findings in Section~\ref{sec:conclusion}.

\section{Observations}
\label{sec:observations}

\subsection{Target selection}

Our targets were selected from the catalogue of \cite{Sewilo2013}, who identified 26 high-reliability and 37 possible YSO candidates in NGC~346 using colour-magnitude cuts and fits to spectral energy distributions (SEDs) with YSO models \citep{Robitaille2006}. These classifications are based on: {\em Spitzer} IRAC (3.6-8.0 $\mu$m) and MIPS (24 and 70 $\mu$m) data from the SAGE-SMC survey \citep{Gordon2011}, near-infrared (JHK$_{\rm s}$) data from the Infrared Survey Facility (IRSF) Magellanic Clouds Point Source Survey \citep{Kato2007} and the Two Micron All Sky Survey (2MASS; \citealt{Skrutskie2006}), and optical (UBVI) data from the Magellanic Clouds Photometric Survey (MCPS; \citealt{Zaritsky2002}) and OGLE-III \citep{Udalski2008}. \cite{Sewilo2013} define high-reliability YSO candidates as either satisfying multiple YSO colour-criteria or are well-fit with YSO models, whilst possible YSO candidates fulfill at least one colour-magnitude criterion and their environment supports their young nature. 
All of our targets in NGC~346 were also observed in the mid-IR by \cite{Simon2007} using data from the S$^3$MC survey \citep{Bolatto2007}.

The YSO candidates identified by these  galaxy-wide surveys present the first opportunity to study a large sample of embedded YSOs in another galaxy in detail. However,  IFU spectroscopy is required to confirm that these objects are indeed single massive YSOs, or if multiples are detected to conclusively identify the massive YSOs from  other sources in the compact proto-cluster \citep[see e.g.,][]{Carlson2012}. To ensure a sufficiently high signal-to-noise (S/N) ratio and line detection rate in our spectroscopy we made a cut in the $J$-band luminosity of $>$0.08 mJy in the NGC~346 catalogue from \cite{Sewilo2013}. From this list of YSO candidates, 18 were observed with KMOS, with the final target selection defined by the KMOS arm allocation software {\sc karma} \citep{Wegner2008}. 
Priority was given to the high-reliability YSO candidates, with the possible high-mass YSO candidates used as fillers. In total 15 high-reliability and 3 possible YSO candidates were observed. 

\begin{figure}
\centering
\includegraphics[trim=0.1cm 1cm 2cm 1.5cm, clip=true, width=\columnwidth]{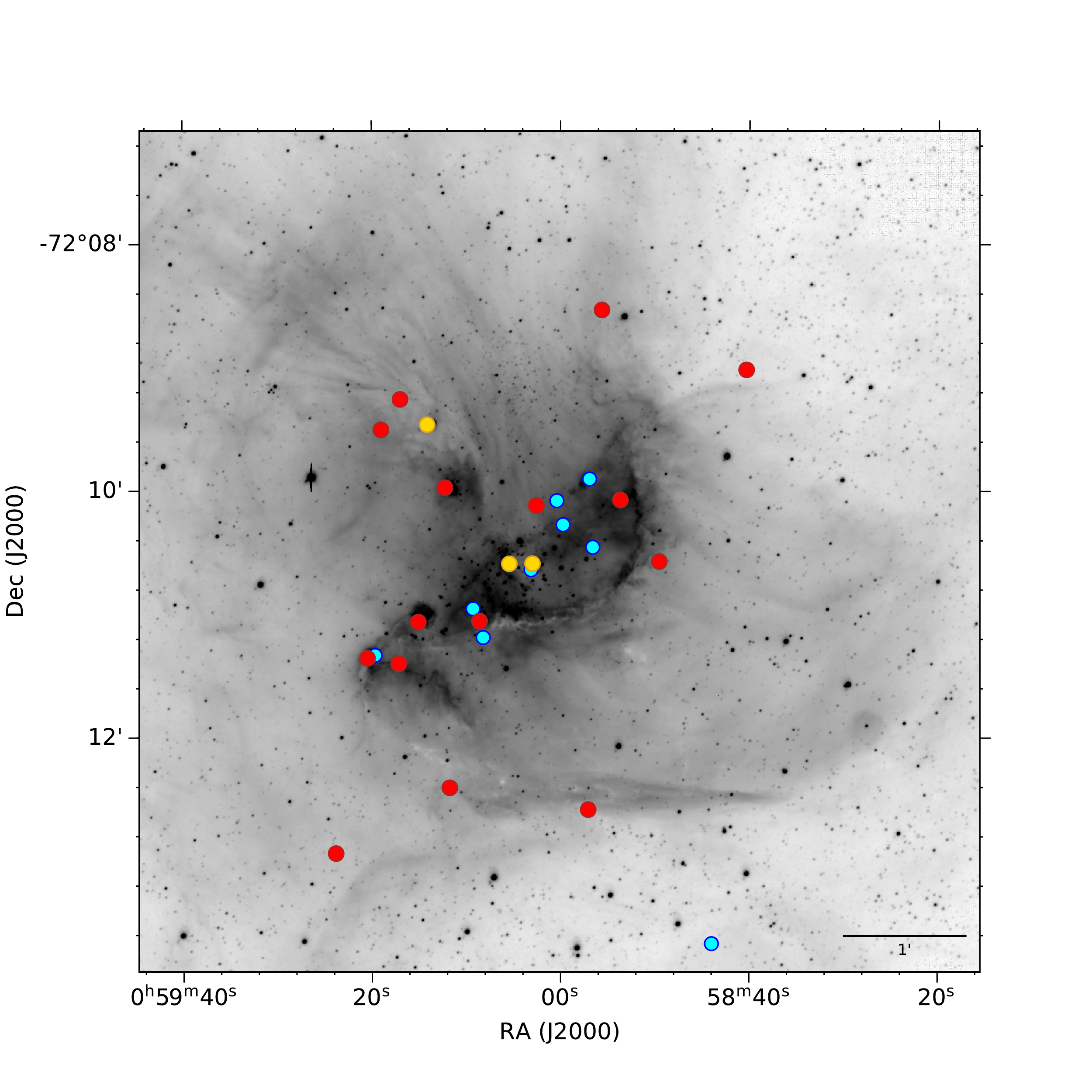}
 \caption{ H$\alpha$ FORS1 map of NGC~346 with north pointing up and east to the left. The location of the high-reliability YSO candidates from \protect\cite{Sewilo2013} observed with KMOS are shown in red, with the remaining unobserved high-reliability candidates shown in cyan. Three `possible' YSO candidates marked with gold symbols from \citeauthor{Sewilo2013} were also observed using otherwise unallocated KMOS arms.}
  \label{fig:NGC346_obs}
  \end{figure}

A summary of the observed targets and their physical properties is given in Table~\ref{tab:obs_summary}, including masses and luminosities from \citet{Sewilo2013} or, in the case of Y519, Y545 and Y548, from \citet{Simon2007}, which were derived by fitting their SED with YSO models from \citet{Robitaille2006}. Their locations are shown in Figure~\ref{fig:NGC346_obs} on a H$\alpha$ image obtained with the VLT FORS1 instrument on 1999 August (PI: Tolstoy; 063.N-0560).
The majority of our candidate protostars are concentrated along the bright dust ridges/filaments in the main \Hii~region. 
 In three instances no values were provided in the literature for the source luminosity. In these cases we derive an estimate for the bolometric luminosity via a simple trapezoidal integration of the SED. This method has been used by \citet{Woods2011}, \citet{Ruffle2015} and \citet{Jones2017b} for sources at all evolutionary stages including YSOs in both the LMC and SMC.

\subsection{KMOS observations}

KMOS is an integral-field spectrograph with 24 deployable integral-field units (IFUs) that can be deployed across a 7\farcm2 field on the sky. Each IFU has a field-of-view of 2\farcs8~$\times$~2\farcs8, sampled by 14~$\times$~14 spatial pixels (spaxels), i.e. each spaxel has an angular size of 0\farcs2. 
Medium-resolution (R~$=$~3000-4000) YJ, H and K band spectra of the 18 candidate YSOs were obtained with KMOS over two nights on 2018 Sep 28-29 (PI: Jones; 0101.C-0612(A)), with total science integrations of 8400s, 1620s  and 3480s in the YJ, H and K bands. The observations were taken in good conditions, with a typical seeing of $<$0\farcs8.
Offset sky frames were obtained using a standard object-sky-object pattern, with each integration dithered to improve bad-pixel rejection from the final extracted spectra. Standard stars were also observed throughout both nights for telluric correction.

\begin{table*}
\centering
\caption{Summary of VLT-KMOS targets in NGC 346.  Luminosities, temperatures and mass estimates are from \citet{Sewilo2013} from SED model fits \citep[except for Y519, 545 and Y548 which are from][]{Simon2007}. If a target was poorly fit by the SED models the bolometric luminosity was estimated from the SED over the range of the available photometry.}
\label{tab:obs_summary}
\begin{tabular}{@{}cccccccccccc@{}}
\hline
\hline
ID	&	RA	&	Dec	&	J	&	H	&	K	& L$_{\rm star}$	&	T$_{\rm star}$	&	M$_{\rm star}$	&	M$_{\rm env}$	&	M$_{\rm disc}$	&	Evol.	\\
	&	[deg]	&	[deg]	&	[mag]	&	[mag]	&	[mag]		&	[\Lsol]   &	[K]	    &	[\Msun]	    &	[\Msun]	    &	[\Msun]	    &  Stage          \\
\hline
Y519	&	14.66801	&	$-$72.15031	&	19.47	&	19.08	&	17.32	&	\pp1590	&   \ldots	&	6.0 	&	\ldots	& \ldots		&	II	\\
Y523	&	14.70647	&	$-$72.17622	&	17.77	&	16.69	&	16.68	&	13400	&	\pp6170	&	15.5	&	3280	&	0.0445	&	I	\\
Y524	&	14.72361	&	$-$72.16790	&	15.99	&	15.61	&	14.38	&	\pp4190	&	\pp5420	&	11.3	&	1250	&	1.6	&	I	\\
Y525	&	14.73176	&	$-$72.14223	&	17.53	&	16.56	&	15.74	&	\pp8580	&	27000	&	11.0	&	1.40E$-$06	&	0.265	&	II	\\
Y528	&	14.73779	&	$-$72.20974	&	19.70	&	18.32	&	17.42	&	\pp5910	&	25500	&	9.84	&	0.000613	&	4.93E$-$07	&	III	\\
Y532	&	14.76068	&	$-$72.16866	&	16.08	&	15.30	&	14.24	& \pp5180	&	12700	&	9.71	&	198	&	0.0356	&	I	\\
Y533	&	14.76242	&	$-$72.17651	&	14.93	&	14.90	&	14.97	&	\pp4940	&	\ldots	&	7.8   	&	\ldots	&	\ldots	&	\ldots	\\
Y535	&	14.77266	&	$-$72.17653	&	14.49	&	13.44	&	12.13	&	27200	&	\ldots	&	\ldots	&	\ldots	&	\ldots	&\ldots		\\
Y538	&	14.78580	&	$-$72.18427	&	17.19	&	16.33	&	15.88	&	\pp6000	&	18800	&	8.44	&	13.4	&	0.00174	&	I	\\
Y543	&	14.79897	&	$-$72.20679	&	18.90	&	18.63	&	17.12	&	40400	&	\pp5600	&	21.4	&	6250	&	0	&	I	\\
Y544	&	14.80103	&	$-$72.16624	&	15.15	&	14.57	&	13.63	&	\pp3240	&	\pp7100	&	10.1	&	21.3	&	1.92	&	I	\\ 
Y545	&	14.80886	&	$-$72.15774	&	15.43	&	15.29	&	14.78	&	\pp8380	&	\ldots	&  11.0  	&	\ldots	&	\ldots	&	II	\\  
Y547	&	14.81283	&	$-$72.18440	&	15.17	&	14.81	&	13.34	&	\pp3580	&	\pp8050	&	10.5	&	25	&	0.0111	&	I	\\
Y548	&	14.82079	&	$-$72.15431	&	16.81	&	15.99	&	15.34	&	\pp\pp653	&\ldots		&	5.5	&	\ldots	&	\ldots	& II	\\  
Y549	&	14.82146	&	$-$72.19002	&	17.09	&	16.41	&	15.82	&	14200	&	29400	&	13.1	&	62.5	&	0.254	&	II	\\
Y551	&	14.82924	&	$-$72.15842	&	19.18	&	18.44	&	14.83	&	\pp\pp392	&	\pp8840	&	5.13	&	39.6	&	0.0249	&	I	\\
Y553	&	14.83529	&	$-$72.18928	&	16.66	&	16.16	&	15.47	&	\pp6150	&	\ldots	&\ldots		&\ldots		&	\ldots	&\ldots		\\
Y556	&	14.84927	&	$-$72.21566	&	17.50	&	17.22	&	16.68	&	\pp5910	&	25500	&	9.84	&	0.000613	&	4.93E$-$07	&	III	\\
\hline
\end{tabular}
\end{table*}

\subsection{Data reduction}
\label{sec:datareduction}

The KMOS IFU datacubes were reduced using the ESO pipeline (version 2.6.3) provided as part of the ESO Reflex automated data reduction environment \citep{Freudling2013, Davies2013_KMOS}. This flat fields, wavelength calibrates, and telluric corrects the raw data, including removing atmospheric absorption using {\sc molecfit} \citep{Kausch2015, Smette2015}. Sky subtraction using a classic nod-to-sky sequence was also performed using the KMOS pipeline. Standard star observations were reduced in an identical manner to the science data; these are then used to flux calibrate the science datacubes. 

To determine the precise location and number of sources in each of our data cubes we use the python {\sc DAOStarFinder} function \citep{photutils, Stetson1987} to detect point sources in the collapsed continuum image. Initially we set the detection threshold at the 3-sigma noise level above the median absolute deviation to determine if the {\em Spitzer} YSO candidate is a member of a compact cluster, and then increased this threshold to the 5-sigma level to determine a source list for spectral extraction.
The full list of sources, their positions and the bands they were detected in are given in Table~\ref{tab:sourceALLPositions}.
As found from higher angular-resolution follow-up of other YSOs in the Clouds identified with {\em Spitzer} \citep[e.g,][]{Chen2009}, many of our {\em Spitzer} targets ($\sim$50\%) are resolved into multiple sources in the KMOS data-cubes. This is not surprising given the seeing of our observations ($<$0\farcs8) compared to the angular resolution of IRAC (1\farcs8). Indeed, we identified multiple sources (using the 3-sigma cut) in half (9/18) of our targets. 

One-dimensional science spectra were obtained using aperture extraction for each point-source observed in the data cubes.  A circular aperture with a 2-pixel radius was used, centred on the peak of the source to ensure the signal is dominated by stellar emission and to minimise the background contribution. A larger annulus scaled to the same area of the aperture was used to measure and subtract the local background. 
For each extracted spectrum we estimated its S/N from a set line-free region for each band using the {\sc snr} function from the {\sc specutils} python package \citep{specutils}. Only spectra with S/N $>$10 were retained for further analysis. 
In total we extracted 26 YJ, H or K band spectra for 12 sources to which we applied a seven-pixel boxcar smoothing filter.

\section{Results}
\label{sec:results}

With the KMOS IFU we observed 18 high-mass YSO candidates identified from {\em Spitzer} data. In these datacubes multiple components can be present. In this section, we analyze the 15 sources in the datacubes that have a 5-$\sigma$ detection from the collapsed continuum images and thus extracted spectra.
Of these 12 sources have spectra with a signal-to-noise ratio (S/N $>$10) sufficient for reliable line detection and measurement.

\subsection{Continuum emission}
\label{sec:Continuumemission}

\noindent Continuum maps for the YJ-, H- and K-band data-cubes are shown in Figures~\ref{fig:contEmission_YJ1}--\ref{fig:contEmission_K2}, which were generated by summing the spaxels along the wavelength axis of the cubes. With such images we can examine the immediate surroundings of the IRAC-identified YSOs, and identify multiple faint sources. 

In our sample of 18 YSO candidates identified by \citet{Sewilo2013}, Y544 and Y551 are clearly resolved into two sources over multiple bands. Seven other candidates also show evidence of multiplicity  (see Table~\ref{tab:sourceALLPositions}), although the companion sources are generally faint and in most cases we are unable to reliably measure their properties and astrometric positions. 
The RA and Dec (J2000) of each extracted source (with at least a 5-$\sigma$ detection in the continuum data) are given in Table~\ref{tab:specPositions}. The positions are determined from the KMOS K-band continuum images. Multiple sources within the same IFU are denoted by "A" or "B" in their name. 

In each field we also searched for spatially-extended line emission beyond the continuum sources. No significant Br$\gamma$ emission or H$_2$ emission was detected in the moment zero maps outside of the compact source region. 
This lack of notable extended emission is likely due to a difference in the spatial resolution between KMOS and high spatial resolution adaptive-optic IFU instruments like VLT/SINFONI (Spectrograph for INtegral Field Observations in the Near Infrared). With SINFONI (spatial resolution of 0.1--0.2 arcsec), \citet{Ward2017} detect spatially extended Br$\gamma$ or H$_2$ emission towards approximately half their YSO sample in the SMC. This is a factor of 6 better sampled than our seeing-limited KMOS data and may explain the difference.

\begin{table}
\centering
\caption{RA and Dec (J2000) for each source with a $>$5$\sigma$ detection in the KMOS K-band images. Those with extracted KMOS spectra with S/N $>$10 in the YJ, H and K regions are 
indicated in the final three columns.}
\label{tab:specPositions}
\begin{tabular}{@{}cccccc@{}}
\hline
\hline
ID	&	RA	&	Dec	&	\multicolumn{3}{c}{S/N $>$10 spectra}\\
	&	[deg]	&	[deg]	&	YJ	&	H &	K	\\
\hline
Y525A	&	14.73175	&	$-$72.14212	&	\checkmark	&	\ldots	&   \ldots	\\	
Y532A	&	14.76073	&	$-$72.16867	&	\checkmark	&	\checkmark	&	\checkmark	\\	
Y533A	&	14.76206	&	$-$72.17640	&	\checkmark	&	\checkmark	&	\checkmark	\\	
Y535A	&	14.77241	&	$-$72.17649	&	\checkmark	&	\checkmark	&	\checkmark	\\	
Y538A	&	14.78541	&	$-$72.18408	&	\ldots	&	\ldots	&	\checkmark	\\	
Y544A	&	14.80079	&	$-$72.16629	&	\checkmark	&	\checkmark	&	\checkmark	\\	
Y544B	&	14.80131	&	$-$72.16624	&	\ldots	&	\ldots	&	\checkmark	\\	
Y545A	&	14.80875	&	$-$72.15780	&	\checkmark	&	\checkmark	&	\checkmark	\\	
Y548A	&	14.82105	&	$-$72.15406	&	\checkmark	&	\ldots	&	\checkmark	\\	
Y549A	&	14.82122	&	$-$72.18994	&	\ldots	&	\ldots	&	\checkmark	\\	
Y553A	&	14.83569	&	$-$72.18915	&	\checkmark	&	\ldots	&	\checkmark \\	
Y556A	&	14.84910	&	$-$72.21564	&	\checkmark	&	\checkmark	&	\checkmark	\\	
\hline
Y524A	&	14.72315	&	$-$72.16779	&	\ldots	&	\ldots	&	\ldots	\\
Y528A	&	14.73675	&	$-$72.20973	&	\ldots	&	\ldots	&	\ldots	\\
Y543A	&	14.79879	&	$-$72.20650	&	\ldots	&	\ldots	&	\ldots	\\
Y551A	&	14.83005	&	$-$72.15836	&	\ldots	&	\ldots	&	\ldots	\\
Y551B	&	14.82900	&	$-$72.15835	&	\ldots	&	\ldots	&	\ldots	\\
\hline
\end{tabular}
\end{table}

\subsection{Properties of the spectra}

\begin{figure*}
  \centering
   \includegraphics[trim=0mm 0mm 0mm 0mm,angle=0,scale=0.76]{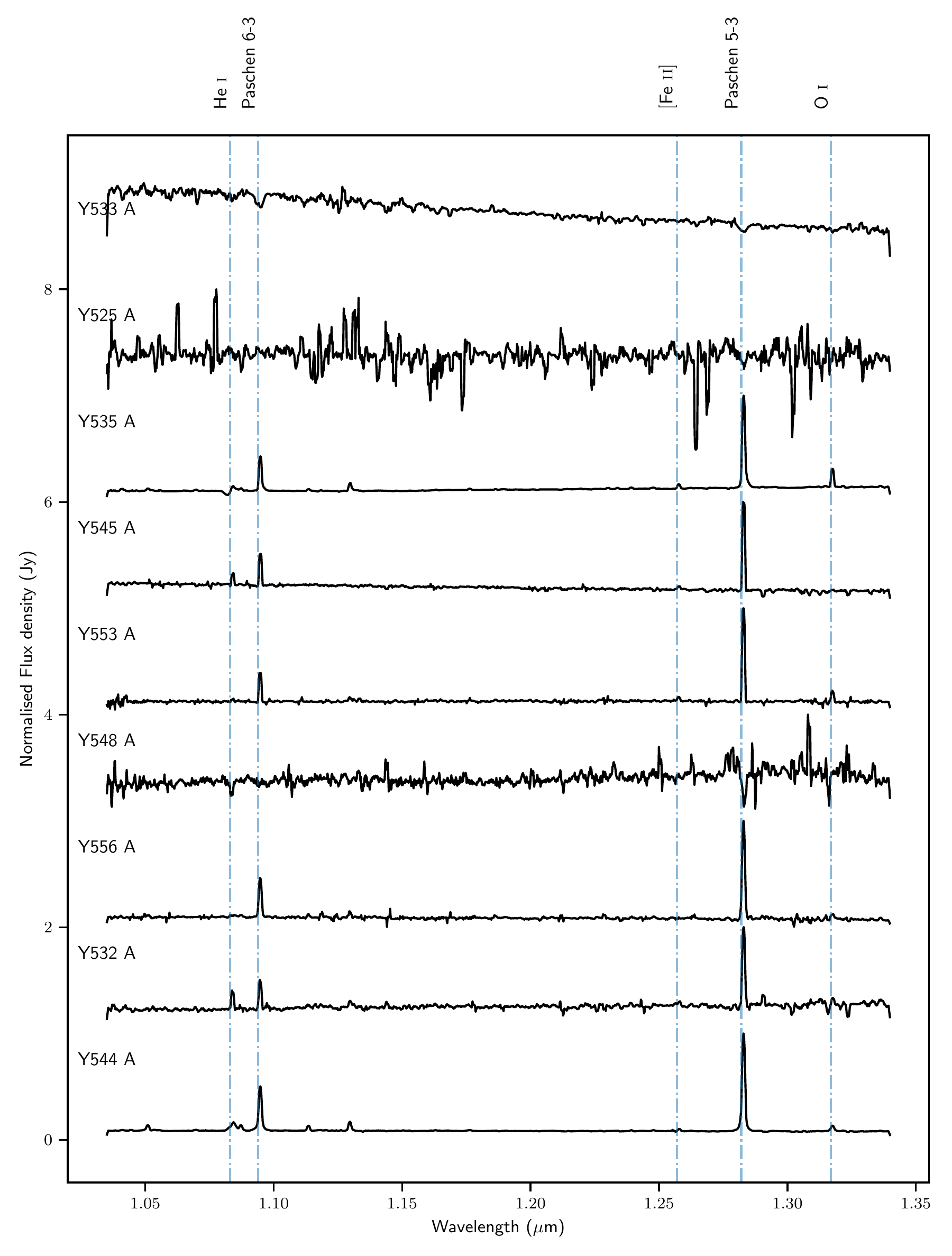}
\caption{Normalised YJ-Band KMOS spectra with S/N $>$ 10 for our YSO candidates. Key emission-lines are marked by the vertical dash-dot lines. For display purposes the flux of each object has been shifted by a continuum unit so the spectra do not overlap. } 
\label{fig:spec_YJ}
\end{figure*}

\begin{figure*}
  \centering
   \includegraphics[trim=0mm 0mm 0mm 0mm,angle=0,scale=0.76]{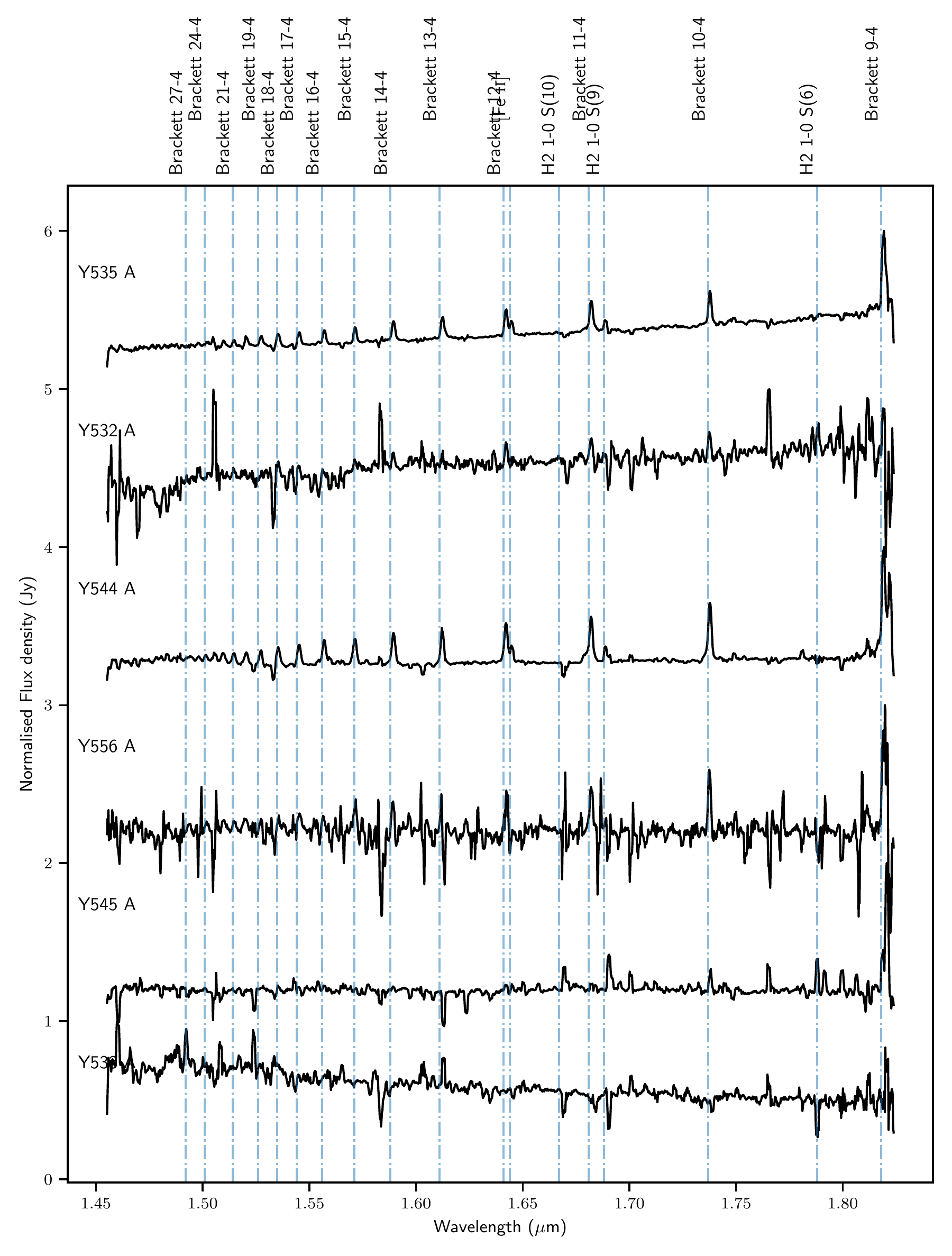}
\caption{Normalised H-Band KMOS spectra with S/N $>$ 10 for our YSO candidates. Key emission-lines are marked by the vertical dash-dot lines. The spectra between each source are offset by a factor of one.} 
\label{fig:spec_H}
\end{figure*}

\begin{figure*}
  \centering
   \includegraphics[trim=0mm 0mm 0mm 0mm,angle=0,scale=0.76]{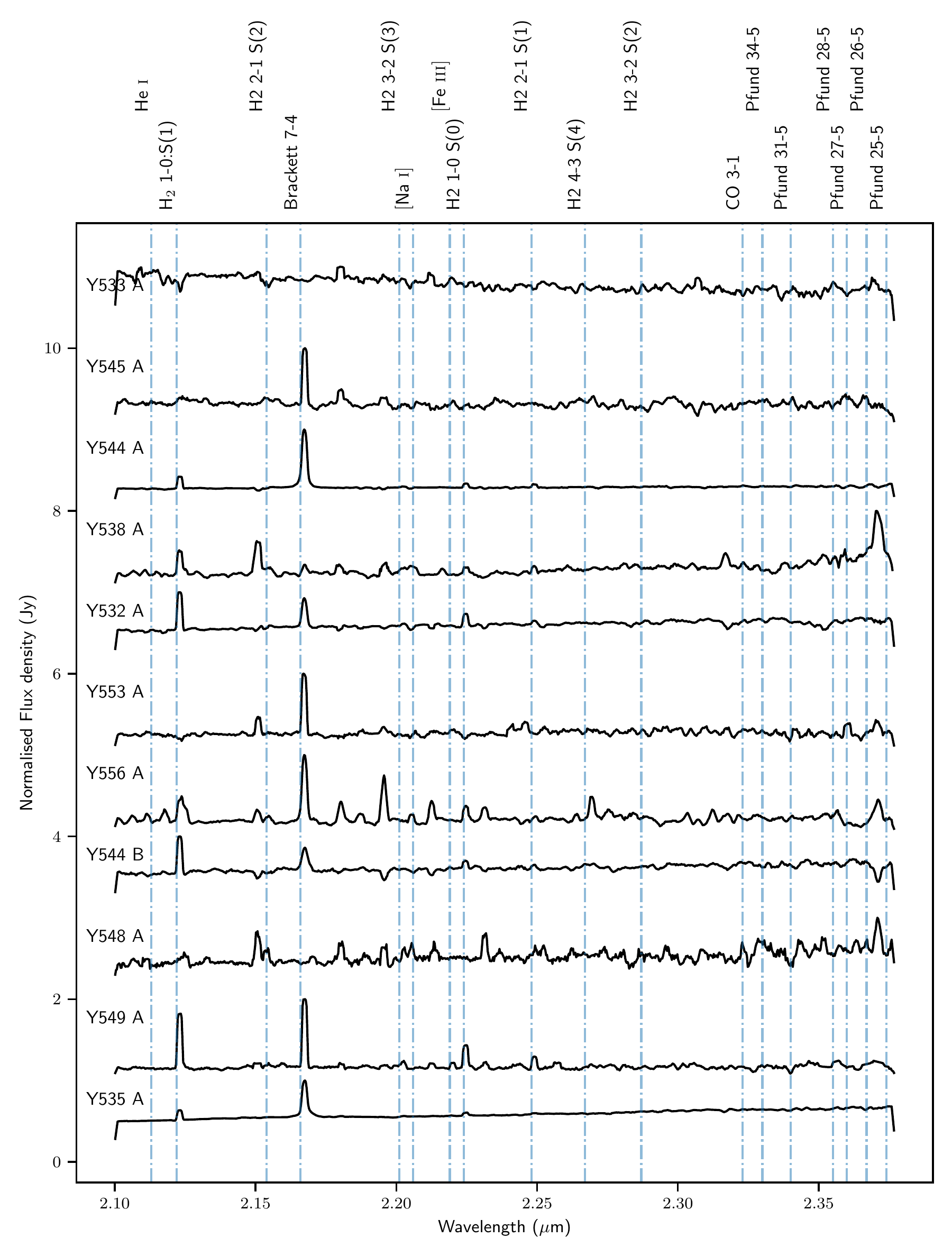}
\caption{Normalised K-Band KMOS spectra with S/N $>$  10 for our YSO candidates. Key emission-lines are marked by the vertical dash-dot lines. The spectra between each source are offset by a factor of one.} 
\label{fig:spec_K}
\end{figure*}

The KMOS spectra of the 12 sources (with S/N $>$10) provide (partial) coverage over $\sim$0.8 - 2.5~$\mu$m. Six sources were detected in all three KMOS bands, three have only S/N $>$10 in the K-band and one (Y525A) has only S/N $>$10 in the YJ-band. The nine YJ band, six H band, and eleven K band spectra of the twelve YSO candidates are shown in Figures~\ref{fig:spec_YJ} -- \ref{fig:spec_K} (and the bands in which each source is detected are summarised in Table~\ref{tab:specPositions}).


Massive YSOs typically have hydrogen recombination lines (e.g. Br$\gamma$) and H$_2$ lines in their spectra, H~{\sc ii} regions have strong H~{\sc i} emission, and O-type spectra can be identified from the ratio of He~{\sc i} to He~{\sc ii} absorption lines. To assess the evolutionary status of each source the KMOS spectra were therefore classified  where possible as either YSOs or early-type stars according to the shape of their continuum and the emission and/or absorption features present in their spectra.

\begin{table}
\begin{minipage}{\columnwidth}
\caption{Common emission lines in the KMOS spectra ($>$3$\sigma$ detections) and their detection frequency within our sample. Note that we do not have sufficiently good spectra (i.e. S/N $>$10) for each source in each KMOS band; to account for those with limited spectral coverage detections are presented relative to the total number of spectra in each band.}
\label{tab:linedetections}
\centering
\begin{tabular}{lccc}
\hline
\multicolumn{2}{c}{Emission Feature} & Wavelength  & Number of   \\
         &        &  ($\umu$m)  & Sources     \\
\hline
Brackett 7-4    &   Br$\gamma$     & 2.166 &  9 of 11  \\      
H$_2$ 1-0:S(1)  &                  & 2.128 &  7 of 11  \\     
Paschen 5-3     &   Pa$\beta$      & 1.282 &  6 of 9   \\  
Paschen 6-3     &   Pa$\gamma$     & 1.094 &  6 of 9   \\ 
H$_2$ 1-0 S(0)  &                  & 2.224 &  5 of 11  \\     
Brackett 10-4   &                  & 1.737 &  5 of 6   \\     
Brackett 11-4   &                  & 1.681 &  4 of 6    \\     
\multicolumn{2}{l}{Brackett 12-4  / [Fe~{\sc ii}]}    & 1.641 &  4  of 6   \\     
\multicolumn{2}{l}{Fe~{\sc ii} /H$_2$ 1-0 S(9)}      & 1.689 &  4 of 6    \\    
$[$Fe~{\sc ii}$]$               &  & 1.257 &  4 of 9   \\     
He~{\sc i}                      &  & 1.083 &  4 of 9    \\   
\hline
\end{tabular}
\end{minipage}
\end{table}

\subsubsection{Observed emission lines}
\label{sec:EmissionFeatures}

The KMOS spectra show a rich variety of emission lines from atomic and molecular species. In the YJ-band strong Paschen series transitions from Pa$\gamma$ (Pa 6-3, 1.094 $\mu$m) and Pa$\beta$ (Pa 5-3, 1.282$\mu$m) dominate the spectra. 
Hydrogen recombination lines from the Brackett series are prevalent in the H-band, with the strongest (secure) detection being Br 10-4 at 1.737 $\mu$m (the Br 9-4 line is stronger but lies near the edge of the observed range, where line measurements have larger uncertainties due to noise).A number of  H$_2$ lines are also present, as is emission from shocked [Fe {\sc ii}] at 1.644 $\mu$m. Finally in the K-band, the most prominent feature is from Br$\gamma$ (Br 7-4) emission at 2.16~$\mu$m, with other atomic hydrogen lines from the weaker Pfund series also present. 
Table~\ref{tab:linedetections} summarises the observed emission lines (with at least 3$\sigma$ detections) and their detection frequency within our sample. 
A full listing of lines detected in each source is given in Table~\ref{tab:linefluxes}. Although there are some weak OH-line residuals after sky subtraction in some of our spectra, we note that our study is helped by the fact that the wavelengths of the strongest sky emission lines do not coincide with the primary diagnostic features for YSOs \citep{Rousselot2000}.

\subsubsection{Emission-line strengths}
\label{sec:EmissionLineStrengths}

For each emission line in the extracted 1D spectra we measured the line flux using a Gaussian profile. 
Spectral lines were initially identified using the {\sc find\_lines\_threshold} algorithm from the {\sc specutils}\footnote{\protect\url{https://specutils.readthedocs.io/}} python package. Only lines with a $>3\sigma$ detection above the continuum were considered. A local continuum, represented by a third-order polynomial, was then fit to each line region via a Levenberg-Markwardt least-square algorithm and subtracted from the spectra. 
Finally, the emission-line strength, position and full width at half-maximum (FWHM) were measured for each line by fitting a Gaussian function to the line profile in the continuum-subtracted spectra. 
When a line was not detected a 3$\sigma$ upper limit was estimated using $3 \times F_{\rm noise} \times \Delta\lambda$, where $F_{\rm noise}$ is the continuum rms noise in the line region and $\Delta\lambda$ is the expected line width, which is assumed to be  0.0014 $\mu$m. 
Table~\ref{tab:linefluxes} presents the measured emission line fluxes, these have not been corrected for extinction.

\subsection{Extinction}
\label{sec:extinction}

Extinction can be calculated using line ratios.  From the observed Pa$\beta$/Br$\gamma$ line ratio we compute the relative extinction between the wavelengths using:
\begin{equation}  
A_{\rm rel} = 2.5 \times \log \left( \frac{ (Pa\beta / Br\gamma)_{\rm obs} }{ (Pa\beta / Br\gamma)_{\rm exp} } \right)
\end{equation}   
where `obs' and `exp' are the observed and expected flux ratios.  Assuming Case B recombination which is valid for $100 < n_e < 10^4$~cm$^{-3}$ and $5000 < T_e < 10^4$~K 
\citep{Storey1995}, the intrinsic flux ratio should be Pa$\beta$/Br$\gamma = 5.75 \pm 0.15$. 
The relationship between the relative extinction due to dust attenuation and $E(B-V)$  is given by: 
\begin{equation} \label{eq:ce}
E(B-V) =\frac{2.5}{k(\lambda_{Br\gamma})-k(\lambda_{Pa\beta})} \log_{10} \Bigg[\frac{(Pa\beta / Br\gamma)_{obs}}{(Pa\beta / Br\gamma)_{int}}\Bigg]
\end{equation}
where $k(\lambda_{Br\gamma})$ and $k(\lambda_{Pa\beta})$ are the reddening curves evaluated at the $Br\gamma$ and $Pa\beta$ wavelengths, respectively. 
Here we use the extinction curves for the SMC derived by \citet{Gordon2003}, and adopt $R_V = 2.74$ which is the average value for the SMC bar. The choice of $R_V$ affects the shape of the extinction curve, and literature values for the SMC range between 2.05 -- 3.30, with higher values resulting in increased  $A_V$ values. 
The computed $E(B-V)$ and $A_V$ values are given in Table~\ref{tab:extinction}. Robust estimates for either (or both) lines were not possible in three sources with YJ and K band spectra, namely: Y544B, Y548A, and Y549A. This leaves the 6 sources with Pa$\beta$ values listed in Table~\ref{tab:linedetections} with extinction measurements listed in Table~\ref{tab:extinction}.

Extinction corrections were calculated for each emission line individually. The $A_{\lambda}$ extinction in magnitudes at the wavelength $\lambda$ is given by:
\begin{equation} \label{eq:intrinsic1}
A_{\lambda}=k(\lambda)E(B-V).
\end{equation}
The intrinsic luminosity of each line, $L_{\rm int}$, is then obtained from:
\begin{equation} \label{eq:intrinsic0}
L_{\rm int}(\lambda)=L_{\rm obs}(\lambda)10^{0.4 A_{\lambda}}, 
\end{equation}
where the observed line luminosities $L_{\rm obs}$ are calculated assuming a distance to the SMC of 62.44 $\pm$ 0.47 kpc \citep{deGrijs2015}. For more details on relating hydrogen recombination lines to attenuation due to dust see \eg \citet{Calzetti2000} and \citet{Momcheva2013}.

The median extinction for our sources is $A_V$ = 1.54 $\pm$ 0.48. This is comparable to the values \citet{Ward2017} obtained in the optical emission for YSOs in the SMC, but significantly lower than their mean extinction values calculated from lines solely in the K-band. This suggests that Pa$\beta$/Br$\gamma$ ratio samples a shallower region of the YSO environment compared to estimates using the H$_2$ Q-branch.
The visual extinctions towards our sample are also relatively low compared to estimates from broadband photometry for massive YSOs in the Galaxy \citep{Cooper2013} and the LMC \citep{Ward2016}. 
Compared to the most extreme sources in these samples our spectra have relatively flat continua and do not show strong dust extinction; consistent with lower dust-to-gas ratios in the SMC. 
Furthermore as the hydrogen recombination lines arise from optically thin regions of a YSO, they do not provide a complete picture of extinction in that source and the true value for extinction is probably much higher, thus rendering direct comparisons to the broadband estimates invalid. 

H$_2$ Q-branch lines together with S-transition lines can also be used to evaluate $A_V$ \citep[e.g.][]{Davis2011}. However, the Q-branch lines occupy a region of the K-band spectrum which is particularly noisy, preventing reliable line ratios to estimate extinction for our sample using this method. Likewise, estimates of $A_V$ via [Fe{\sc ii}] transitions in the J and H-bands are limited by the potential blend of the Br~10 line and [Fe{\sc ii}] at 1.64$\mu$m.

\begin{table}
\caption[Extinction]{Extinction estimates for sources with robust determinations of the Pa$\beta$ and Br$\gamma$ line intensities.}
\centering
\begin{tabular}{lccc}
\hline\hline
ID           & Pa$\beta$/Br$\gamma$    &$E(B-V)$        & $A_V$ \\
 & & [mag] & [mag] \\
\hline 
Y532A	&	2.044 $\pm$ 0.336	&	1.28 $\pm$ 0.21	&	3.51 $\pm$ 0.92	\\	
Y535A	&	3.307 $\pm$ 0.137	&	0.68 $\pm$ 0.06	&	1.86 $\pm$ 0.16	\\	
Y544A	&	4.146 $\pm$ 0.375	&	0.40 $\pm$ 0.12	&	1.10 $\pm$ 0.33	\\	
Y545A	&	5.596 $\pm$ 0.258	&	0.03 $\pm$ 0.06 	&	0.08 $\pm$ 0.16	\\	
Y553A	&	3.370 $\pm$ 0.156	&	0.66 $\pm$ 0.07	&	1.81 $\pm$ 0.19	\\	
Y556A	&	4.389 $\pm$ 0.145	&	0.33 $\pm$ 0.05	&	0.90 $\pm$ 0.14	\\	
\hline
\end{tabular} 
\label{tab:extinction}
\end{table}


\subsection{Stellar radial velocities}
\label{sec:radialVelocities}

Radial velocities ($v_{\rm r}$) of our targets were measured from the Doppler shifts of the \brg, H$_2$ 1-0:S(1), Pa$\gamma$ and Pa$\beta$ lines. 
The line centres were measured by Gaussian profile fits to the emission lines detected with strengths $>$3$\sigma$ above the continuum.  Mean velocities and standard deviations ($\delta v_{\rm r}$) of the individual measurements are given for the eight sources where estimates were possible in Table~\ref{tab:radialVelocity}. The average radial velocity for our targets is 178 $\pm$ 3 \kms, in good agreement with the peak velocity of the H\,{\sc i} gas emission in the
direction of NGC~346 \citep[see, e.g.][]{Stan99}. The mean velocity also compares well with the systemic velocities of massive OBA-type stars in this region \citep[e.g.][]{Evans2006,Evans2008,Dufton2020}, which are notably offset by $\sim$20 \kms\ compared to older stellar populations in the SMC \citep[e.g.][]{HZ2006}.

\begin{table}
\caption[Velocity]{Mean radial velocity ($v_{\rm r}$) estimates for each source.}
\centering
\begin{tabular}{lcc}
\hline\hline
ID  & $v_{\rm r}$ & $\delta v_{\rm r}$ \\
& [\kms] & [\kms] \\ 
\hline 
Y532A	&	174.34	&	2.58	\\
Y535A	&	181.78	&	0.91	\\
Y544A	&	176.05	&	3.49	\\
Y544B	&	189.38	&	9.73	\\
Y545A	&	187.65	&	1.90    \\
Y549A	&	183.95	&	1.04	\\
Y553A	&	159.46	&	2.27	\\
Y556A	&	171.82	&	1.57	\\
\hline
\end{tabular} 
\label{tab:radialVelocity}
\end{table}


\subsection{Comments on individual sources}
\label{sec:IndividualSources}

We see Br$\gamma$ emission in 9 of the 11 K-band spectra in our sample. This is consistent with massive YSOs that have ionized their gas and is indicative of their youthful nature, confirming the YSO photometric classifications from \cite{Sewilo2013} and \cite{Simon2007}. Of particular note are the three sources with only K-band spectra, as they may be located deep inside dense molecular clouds or are still embedded in in-falling envelopes.

\medskip

\noindent We first briefly discuss the two sources without Br$\gamma$ emission:
\begin{itemize}
    \item {\it Y533A:} There is a small offset in the position, but the brightest source in this cube is a O7 dwarf previously observed spectroscopically by \citet[NGC 346 MPG~396]{Massey1989} and \citet[NGC346-1015]{Dufton2019}. There is a faint companion star nearby in the data cube which is perhaps the source of the large {\em Spitzer} IR excess, giving an offset positional centroid compared to the O7 star. Secure line identifications are difficult in the KMOS data, but Pa$\beta$ and Pa$\gamma$ absorption are present in the YJ-band spectrum, with a hint of He~{\sc i} 1.083~$\mu$m absorption. In the H- and K-band spectra the absence of strong absorption lines from the Brackett series is surprising, but we are perhaps limited by the S/N of the data. There are tentative detections of He~{\sc ii} 1.572 and 1.692~$\mu$m, but these are somewhat suspicious given the absence of Brackett absorption lines, as well as the absence of He~{\sc i} 1.700~$\mu$m \citep[see, e.g.][]{Blum1997,Hanson2005}. In short, the YJ-band spectrum is consistent with Y533A being NGC 346 MPG~396, but it is difficult to say much more with the current near-IR data.
    
    \item {\it Y548A:} Secure line identifications are also challenging for Y548A. There is a more secure detection of He~{\sc i} 1.083~$\mu$m absorption, with (noisy) absorption present at the wavelength of Pa$\beta$, but notably with no corresponding detection of Pa$\gamma$. If the Pa$\beta$ absorption is real, it would suggest a cooler object than Y533A; analogous to an early spectral type, between late O and early B.
    \end{itemize}

\noindent We now discuss three other sources of note:
    \begin{itemize}
    
    \item{\it Y535A:} Identified as NGC 346 MPG 454, Y535A is located in the very central part of the NGC 346 complex. Y535A is an intriguing object. It is among the most luminous sources in the region at both X-ray and near-IR wavelengths \citep{Naze2002, Rubio2018}. Furthermore, it appears to be part of a young star complex, with a mid-infrared spectra  (listed as SMC IRS 18 and PS9 in \citealt{Ruffle2015} and \citealt{Whelan2013}, respectively) dominated by a massive YSO exhibiting silicates in emission \citep{Whelan2013, Ruffle2015, Kraemer2017}.  Representing the most evolved YSO type in their schema (see \citealt{Woods2011} for a in depth description and \citealt{Oliveira2013} for the first adaption of this classification scheme to the SMC) Y535A is considered to be a YSO-4 (Herbig AeBe star) by \citet{Ruffle2015} from its {\em Spitzer} Infrared Spectrograph (IRS) data and is spectroscopically classified as a Be star by \citet{Martayan2010} in their slitless H$\alpha$ survey.  It is likely that multiple nearby sources are contributing to the observed flux that cannot be separated out. 
    This bright compact star cluster is surrounded by many young massive stars.  High-resolution {\em HST} data reveal nine stars within 1$''$ of its extracted position \citep{Sabbi2007}, however these sources are unfortunately too faint to be detected in our KMOS data cubes, where Y535A appears as a single object in the H and K bands, with a possible companion detected at the 3-$\sigma$ level in YJ.

    According to \cite{Rubio2018} Y535A (also known as `Source C') is a Stage I YSO, with a mass of 18 \Msun\ enshrouded by a compact \Hii\ region. Their `Source C'  HK-band (R $\sim$ 1000) spectrum is rich in hydrogen recombination, H$_2$ and He\,{\sc i} emission lines. The detected lines and rising continuum towards longer wavelengths in the data from \citeauthor{Rubio2018} are consistent with our near-IR spectra, which exhibits a P-Cygni profile in He~{\sc i}~1.083 $\mu$m that suggests an outflow. Fluorescent Fe\,{\sc ii} at 1.6878 $\mu$m, which originates in discs \citep{Zorec2007}, is present in both the KMOS data and the HK spectra from \citeauthor{Rubio2018} However, we do not detect extended Br$\gamma$ or H$_2$ emission from the source (see Section~\ref{sec:Continuumemission}). Combined, the results confirm the massive YSO nature of Y535A.
    
    \item {\it Y544A \& B:}  Appearing as single sources in {\em Spitzer} data this target was resolved into multiple objects with the KMOS IFU (and in high-resolution {\em HST} images; \citealt{Hennekemper2008}). This YSO candidate was observed with the {\em Spitzer} IRS \citep[][SMC IRS 21]{Ruffle2015}. SMC IRS 21 was classified as a {\sc YSO-4}: a Herbig Ae/Be type object due to the silicate emission features present \citep{Whelan2013, Ruffle2015}. 
    The position of this YSO candidate coincides with that of NGC 346 MPG 605, a red star with ($B-V$)\,$=$\, 0.25 mag from \citet{Massey1989}, which has also been classified as a Be-type star, NGC~346:KWBBe~448 with a B0 spectral type \citep{Keller1999, Martayan2010}. 
    
    In the KMOS data, both Y544A and Y544B have comparatively flat continua, with no excess dust emission at $\lambda >$ 2.3 $\mu$m. Y544A has spectral detections in all three KMOS bands, however, Y544B only has a K-band spectrum with S/N $>$10, suggesting this is the cooler object and/or with larger extinction than Y544A. Fluorescent Fe\,{\sc ii} and [Fe\,{\sc ii}] lines are observed in Y544A compatible with forbidden disc emission potentially from a Herbig Oe/Be disc. 
\end{itemize}

\section{Discussion}
\label{sec:discussion}

\subsection{H\,{\sc ii} emission} 
\label{sec:H_recombinationLines}

Hydrogen recombination lines (e.g., Br$\gamma$, $n = 7-4$ at 2.166~$\mu$m) are a primary diagnostic for the physical properties and kinematics of ionized regions. 
Our spectra are rich in Paschen, Brackett, and Pfund series emission lines. The Br$\gamma$ line is the strongest  H\,{\sc ii} emission line in the $K$-band associated with YSOs, with a detection frequency of 80\% in our sample. It is commonly used to trace accretion luminosity \citep[e.g.][]{Calvet2004, Mendigutia2011, Cooper2013, Ward2016, Ward2017, Reiter2019} in protostars (especially in sources which are heavily extincted) via the relation: 
\begin{equation}  
\log \frac{L_{\mathrm{acc}}}{L_{\odot}} = (0.91\pm0.27) \times \log \frac{L_{\mathrm{\mathrm{Br}\,\gamma}}}{L_{\odot}} + (3.55 \pm 0.80).
\label{eqn:Lacc}
\end{equation}
Metallicity is not expected to influence this relation, as the number of photons capable of ionising hydrogen (at T $>$ 10$^4$ K) in metal-poor environments is comparable to solar metallicity values \citep{Kudritzki2002}. 

\begin{figure}
  \centering
\includegraphics[trim=0cm 0cm 1cm 1cm, clip=true,width=\columnwidth]{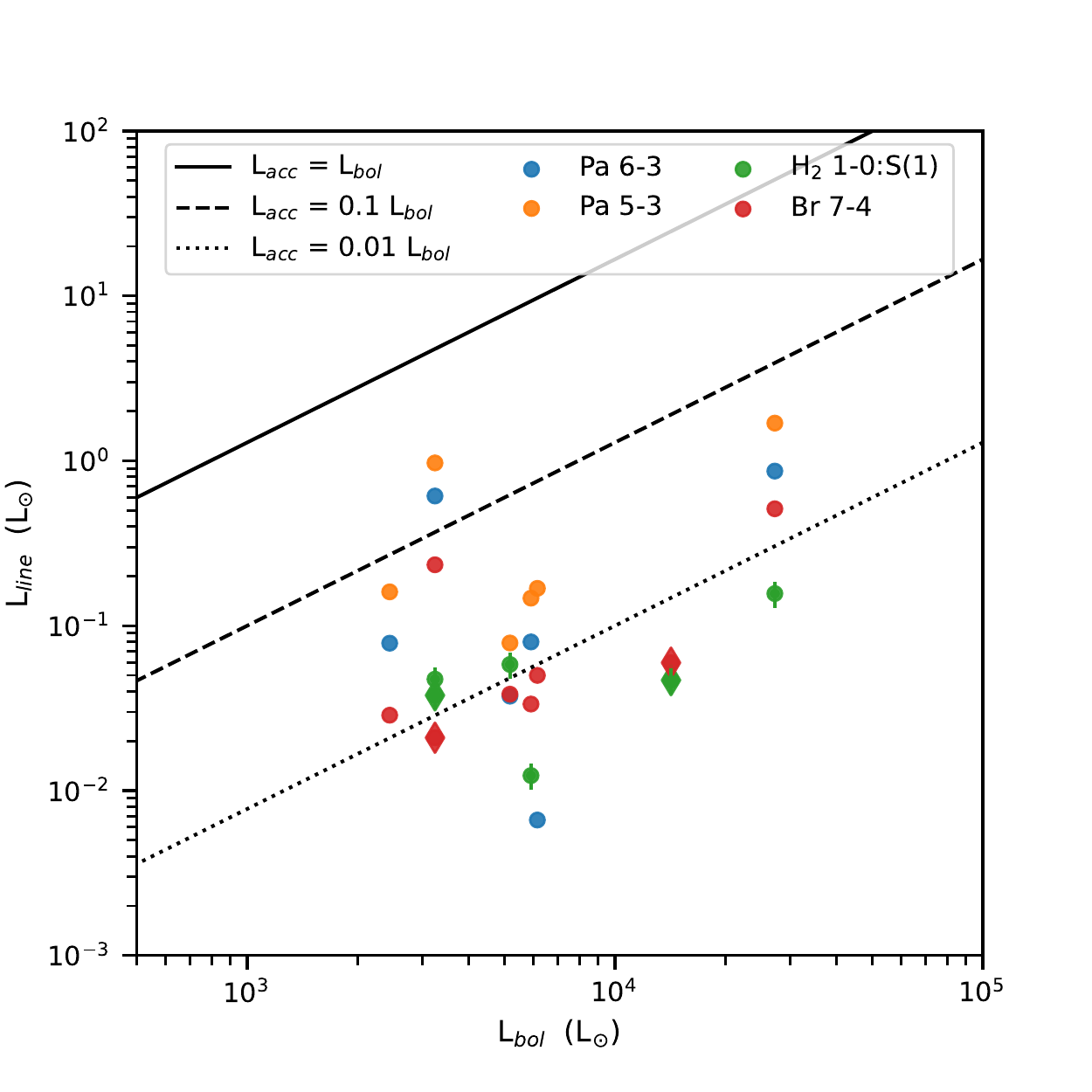}
\caption{Line luminosity vs.~bolometric luminosity (for >3$\sigma$ line detections), in which the line luminosities have been corrected for extinction where possible (circles); uncorrected lines are plotted as diamonds. The error bars for the \brg, Pa$\gamma$ and Pa$\beta$ lines are smaller than the plot symbols. The overplotted lines show the loci of ${\rm{L}}_{{\rm acc}} = {\rm L}_{{\rm bol}}$, $0.1\ {\rm L}_{{\rm bol}}$ and $0.01\ {\rm L}_{{\rm bol}}$ calculated using the relationship from \citet{Mendigutia2011}.}  
\label{fig:linelumVbollum}
\end{figure}

The extinction-corrected luminosities of Br$\gamma$ emission compared to the bolometric luminosity of each source are shown in Figure~\ref{fig:linelumVbollum}. Treating the luminosity from the {\em Spitzer} YSOs as single massive objects is a reasonable assumption as the SED of a compact-(proto)cluster is dominated by its most massive star \citep[e.g.,][]{Bernasconi1996, Chen2009}. For the {\em Spitzer} sources that we resolved into multiple continuum sources, the bolometric luminosity is considered an upper limit. 
We also plot the relation from \citet[][equation~\ref{eqn:Lacc}]{Mendigutia2011} for a range of ${\rm L}_{{\rm bol}}$; most of our objects fall in the $0.01\ {\rm L}_{{\rm bol}} < {\rm{L}}_{{\rm acc}} < 0.1\ {\rm L}_{{\rm bol}}$ range. 

The Pa 5-3 and 6-3 lines are seen in emission in six of the nine objects with YJ spectra. In the H-band there are three strong Brackett series transitions (12-4 at 1.64 $\mu$m, 11-4 at 1.68 $\mu$m and 10-4 at 1.74 $\mu$m) in several sources, with multiple higher-order lines in the Brackett series also present. Pfund series emission is detected in four sources, however this region of the spectrum has a high degree of noise due to the poor atmospheric transmission. Obtaining accurate flux measurements for these lines is therefore challenging and we do not use them in our analysis. 

\begin{figure}
  \centering
    \includegraphics[trim=0.3cm 0cm 0.3cm 0cm, clip=true,width=\columnwidth]{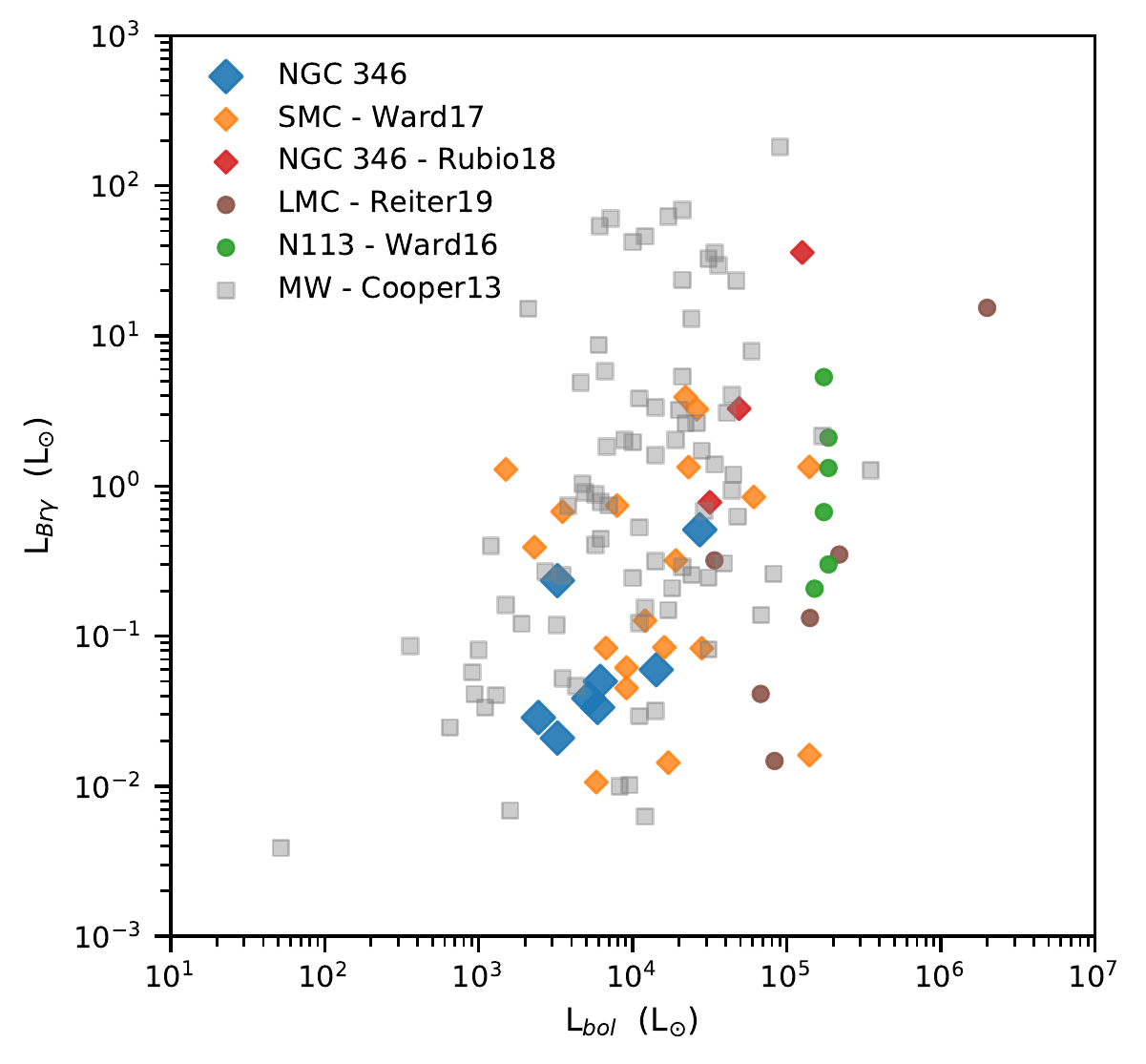}
    \caption{Br$\gamma$ line luminosity vs. bolometric luminosity for our NGC 346 targets (blue diamonds) compared with published samples: three additional objects in NGC 346 \citep[][red diamonds]{Rubio2018}, other YSOs in the SMC \citep[][orange diamonds]{Ward2017}, YSOs in the LMC from \citet[][brown circles]{Reiter2019} and \citet[][green circles]{Ward2016} and YSOs in Galactic high-mass, star-forming regions \citep[][grey squares]{Cooper2013}. For our NGC 346 sample, the error bars for the line luminosity are smaller than the plot symbols. L$_{\rm {bol}}$ values for the Magellanic Clouds should be treated as an upper limit due to the high prevalence of multiplicity. }
  \label{fig:Hi_linelumVbollum} 
\end{figure}

The Br$\gamma$ luminosities for our NGC 346 sample are compared in Figure~\ref{fig:Hi_linelumVbollum} to results for high-mass galactic YSOs discovered in the Red MSX Source (RMS) survey \citep{Cooper2013} and high-mass YSOs in the Magellanic Clouds \citep{Reiter2019,, Ward2016, Ward2017, Rubio2018}\footnote{No line strengths are published for the 10 massive YSOs spectroscopically identified by \citet{vanGelder2020} in the 30 Doradus region of the LMC.}. 
Our results are consistent with those for Galactic YSOs, and are at the fainter end of spectroscopic sources observed in the Magellanic Clouds; especially when compared to the LMC samples which have on the whole higher L$_{\rm {bol}}$ values. Furthermore, the LMC sample includes H72.97$-$69.39 one of the most luminous YSOs in the LMC, which has also been proposed to be a young super star cluster analogous to R136 in the heart of 30 Dor \citep{Ochsendorf2017, Reiter2019}. 
If, as expected, the Br$\gamma$ luminosity is correlated to the mass of the YSO, then our objects would sample the lower-mass end of the massive YSO distribution (i.e. more likely the progenitors of B-type rather than O-type stars).

To compare the SMC and Milky Way samples with luminosities in the range 1--50 $\times10^{3}$  \Lsol~ in a quantitative manner, we divide each sample into three luminosity bins split equally in logarithmic space. In general, the mean and median  Br$\gamma$ line strengths increase as the bolometric luminosity of the source increases, and for a given luminosity range the mean and median Br$\gamma$ luminosity is lower for the SMC sources compared to the Milky Way sources.  However, this trend has considerable scatter and large associated errors, due to the low number of sources in each bin and the large uncertainties associated with the bolometric luminosity. To determine if this effect is real and a consequence of the differences in the metallicity between the populations will require additional high-spatial resolution observations (now possible by {\em JWST}) to place better constraints on both the bolometric luminosity and to reduce potential statistical biases in the samples.

\subsection{H${_2}$} 
\label{sec:shock_h2}

Molecular hydrogen emission lines are commonly observed in dense photodissociation regions and are also a common tracer of shocks from outflows. Large energies (T $>10^3$ K) are required to excite  H$_2$. The detection of H$_2$ emission is prevalent in massive objects in the earliest stage of star formation where bipolar molecular outflows are common. At later stages, the detection rate of H$_2$ drops as the circumstellar envelope dissipates. 
The H$_2$~1-0S(1) line at 2.122 $\mu$m is detected in emission in over 60 \% of our sample. 
Other H$_2$ emission lines are also seen in our sample, although at a reduced rate. A full list of H$_2$ lines detected in each source is given in Table~\ref{tab:linedetections}.  

The H$_2$/Br$\gamma$ ratio can be used to determine excitation conditions in the gas \citep[e.g.,][]{Dale2004, Yeh2015, Reiter2019}. 
Two objects (Y532A and Y544B) have a H$_2$/Br$\gamma$ ratio slightly greater than unity, suggesting shocks excited by protostellar outflows are contributing to the observed flux.
These two objects are likely the youngest sources in our sample. 
For the remainder of the sample H$_2$/Br$\gamma<$1, suggesting photoexcitation is the primary mechanism. 
However, given the number of H$_2$ lines detected towards the objects in our sample, the distribution in line ratios and the uncertainties in these measurements, the true excitation mechanism is likely a combination of both photoexcitation and shock excited processes representing a complex physical environment.

H$_2$ {\em absorption} at 2.122 $\mu$m is tentatively seen towards Y533A, a mid O-type star in NGC 346. H$_2$ absorption lines are relatively weak and are very difficult to observe, requiring unusually hot or radiatively excited gas with sufficiently large column densities \citep{Lacy2017}. This indicates the star Y533A is still surrounded by its natal gas and dust, or is behind the molecular cloud. 


\subsection{[Fe~{\sc ii}] emission} 

[Fe~{\sc ii}] 1.644~$\mu$m emission is often detected in shock-excited gas, including in protostellar jets \citep{Giannini2013} and supernova remnants \citep{Lee2019}.
In \Hii\ regions far-UV radiation can also excite [Fe~{\sc ii}] emission  \citep{Mouri2000}.
In the YJ band, [Fe~II] at 1.257 $\mu$m may be present in Y535A, Y544A, Y545A and Y553A suggesting that these sources have nebular emission. Indeed, \citet{Sewilo2013} noted that Y545 is surrounded by small \Hii~region in the {\em HST} H$\alpha$ data.

Shocked $[$Fe~{\sc ii}$]$ lines at 1.644 $\mu$m are detected in the H-band spectra of two of our sources, Y535A and Y544A. Unfortunately, at this spectral resolution, quantitative analysis of the [Fe~{\sc ii}] is not possible due to blending with the stronger Br~12 line.


\subsection{He\,{\sc i} emission} 

He\,{\sc i} emission at 1.083 $\mu$m is detected in  four of the nine sources with YJ band spectra and,  as noted earlier in Section~\ref{sec:IndividualSources}, displays absorption in Y533A and Y548A. 
To ionize helium hard UV photons are needed, suggesting that only the brightest sources with temperatures $>$ 20,000 K can directly excite the line.  However, collisional excitation of He\,{\sc i} emission from a stellar wind is also possible, explaining why He\,{\sc i} emission is frequently observed  around low- and intermediate-mass T-Tauri and Herbig AeBe stars  \citep[e.g.,][]{Edwards2006,Fischer2008,Reiter2018}. 
The detection of He\,{\sc i} emission in our sample cannot therefore be used to constrain the either the mass or evolutionary stage of these objects \citep{Covey2011, Connelley2014}.
Conversely, He\,{\sc i} can only be seen in absorption when the source has kinetic temperatures above ${\rm T} \ge 10^{4}$ K and electron densities of $n_{e}\ge 10^{8}\ {\rm cm}^{-3}$ \citep{Drew1993}, thus the line must be produced very close to a hot star ( partly the reason for our earlier classifications of Y533A and Y548A as early-type stars).

One source, Y535A, shows a P-Cygni-type profile at He~{\sc i}~1.083~$\mu$m, with blue-shifted absorption and red-shifted emission, characteristic of outflowing gas.  He~{\sc i}~1.083~$\mu$m profiles have been used to trace mass flows around low- and intermediate-mass stars \citep[e.g.][respectively]{Edwards2006,Cauley2014}. 
We note that none of our sample displays profiles with the red-shifted absorption characteristic of accretion flows.


\subsection{Accretion discs: CO bandhead and fluorescent Fe~{\sc ii}}
\label{sec:CObandhead}

CO overtone bandhead transitions at 2.29--2.50 $\mu$m are commonly observed in galactic YSOs \citep[e.g.,][]{Blum2004, Bik2006, Ilee2013}. 
The distinctive CO bandhead emission originates from the inner part of a Keplerian disc of neutral gas which has high densities ($n > 10^{11}$ cm$^{-3}$) and temperatures \citep[2500--5000 K;][]{Chandler1995, Blum2004}. 
Studies of galactic YSOs show that not all sources with an accretion disc exhibit CO bandhead emission, for instance if the accretion rate is too low or the disc cannot self-shield due to low column densities \citep{Cooper2013, Ilee2018b}. It is most often detected in sources with moderate mass accretion rates, 
$\dot{{\rm M}}_{\rm acc} \sim10^{5}$ ${\rm M}_{\odot} \, {\rm yr}^{-1}$.
CO bandhead absorption lines arise in a different region, further from the star in the cooler outer envelope \citep{Davies2010_CO}. 
Strong CO bandhead absorption is common in the spectra of of FU~Ori-like objects -- low-mass young stars undergoing an accretion outburst \citep[e.g.,][]{Connelley2018}.

We do not detect the CO bandhead in emission or absorption in our data. 
In the Magellanic Clouds there have only been four tentative or weak detections of CO bandhead emission, with it seen in absorption in two other cases \citep{Ward2016, Ward2017, Reiter2019, vanGelder2020}. In contrast, 17\% of the Galactic YSOs observed by \citet{Cooper2013} show CO bandhead emission. In higher-resolution data of YSOs drawn from the same Galactic sample (R$\sim$7000 versus R$\sim$500), the detection rate of CO first overtone emission increased to 34\% \citep{Pomohaci2017}.
If we consider all K-band spectroscopic data targeting YSOs in the LMC and SMC (including the KMOS data presented here), the upper-limit to the detection rate for the CO bandhead emission in the Magellanic Clouds is $\sim7\%$. From a total sample of 53 YSOs in the Clouds, only source \#03 in the SMC from \citet{Ward2017} and sources S4, S5-E and S7-A in the 30 Doradus region of the LMC \citep[observed at R $\sim$ 4000 -- 17000;][]{vanGelder2020} have discernible CO bandhead emission. In all instances these detections were described as tentative or weak.  For instance, for the SMC source \#03 of \citet{Ward2017} the 3--1 and 4--2 bandheads have substantial contamination from CO absorption lines and the $v = 2-0$ transition is weak \citep{Ward2017}. This is the only CO bandhead emission that has been detected in the SMC. 
This mounting evidence of significantly lower detection rates in metal-poor environments suggests a physical difference between star-formation mechanisms. We discuss this discrepancy further in Section~\ref{sec:comparison}. 

Fluorescent Fe~{\sc ii} 1.6878 $\mu$m emission is also thought to originate from a dense circumstellar accretion disc \citep{Porter1998, Zorec2007}, in a partially ionized zone between the shielded disc mid-plane and the full ionization zone. 
However, spectra of massive YSOs exhibiting both fluorescent Fe~{\sc ii} and a CO bandhead are rare \citep{Cooper2013}. 
The fluorescent Fe~{\sc ii} transition at 1.6878 $\mu$m occurs when excited atoms/ions (from UV continuum or Lyman series photons) cascade from an upper level of 6.2 eV  back down to the ground state. Fe~{\sc ii} 1.6878 $\mu$m emission lines are detected in both Y544A and Y535A where [Fe~{\sc ii}] and Brackett series emission are also prevalent and there is no evidence for H$_2$ 1-0 S(10) emission. It is possibly present in 
Y545A. However, this line can also be identified as H$_2$ 1-0 S(9) and is the more likely identification in this source.
It is thought that fluorescent Fe~{\sc ii} transition at 1.6878 $\mu$m is more prevalent in massive YSOs \citep{Lumsden2012}, thus a secure detection in two of our sources corroborates their massive status. 


\subsection{Comparison of YSOs in the Magellanic Clouds and Galactic high-mass star-forming regions}
\label{sec:comparison}

It is only within the last decade or so that observing large samples of YSO with near-IR spectroscopy is possible outside our own galaxy.
Including our sample, 53 K-band spectra of moderate to high resolution have been obtained towards YSOs in the Magellanic Clouds \citep{Ward2016,Ward2017,Rubio2018, Reiter2019, vanGelder2020}. These spectra have a diverse range of spectroscopic properties and represent YSOs and unresolved clusters of multiple high-mass stars, across a range of massive YSO evolutionary stages, viewing angles, accretion rates and masses. In Figure~\ref{fig:Hi_linelumVbollum} we compare the Br$\gamma$ line luminosities (where available) for these samples and the sample of Galactic high-mass YSOs from \citet{Cooper2013}. 
In all instances there is a very high detection rate of Br$\gamma$ emission.
Given the range of YSO properties the Br$\gamma$ emission likely originates from several environments including the disc, disc wind and the surrounding low-density ionized gas in \Hii\ regions. 

Another interesting feature in the K-band is the CO bandhead mentioned earlier, which is seen in emission for massive YSOs and is associated with the dust sublimation region in the inner accretion disc \citep{Dullemond2010, Ilee2018b}. 
In the Milky Way, the luminosity of the CO bandhead emission correlates with that of Br$\gamma$ \citep[e.g.][]{Connelley2010, Pomohaci2017}, despite their different kinematic origins \citep{Hsieh2021}. As noted in Section~\ref{sec:CObandhead}, the detection rate of CO bandhead {\em emission} in the Magellanic Clouds is extremely low. 
This raises the intriguing possibility that metallicity substantially affects the physical properties of the accretion disc in massive stars and thus their formation process in metal-poor environments. 

In high-mass stars disc accretion remains poorly understood. Here the detection of CO bandhead emission is thought to be dependent on geometry \citep{Kraus2000, Barbosa2003} or accretion rates \citep{Ilee2018b}. 
Given the diversity of sources and size of both samples, it is unlikely that geometry plays a significant role in the low rate of detection in the Magellanic Clouds. 
Similarly, unless accretion through the inner gaseous disc occurs by a different mechanism (e.g. episodic versus continuous accretion) or occurs at substantially higher or lower rates compared to Galactic sources (which is not substantiated by the observed Br$\gamma$ luminosities; see Figure~\ref{fig:Hi_linelumVbollum}) the range of mass accretion rates (and stellar properties of the embedded YSOs that cover the mass range of 5--30 M$_\odot$) appear comparable for both populations and are therefore unlikely to account for the lack of CO bandhead emission. 

Dust and gas characteristics depend on the metallicity of the host galaxy. This is a more probable explanation for the low CO bandhead detection rate.
In the SMC there is a lower gas-phase CO abundance \citep{Leroy2007SMC}, higher ambient UV radiation field, and lower dust content compared to solar metallicity gas.  
Lower CO abundances and reduced shielding from dust allows the intense UV flux to penetrate further into molecular clouds and circumstellar discs. 
The excitation of the CO bandhead emission requires warm temperatures (T = 2500--5000 K) and high densities ($n> 10^{11}$~cm$^{-3}$) so it is thought to trace the inner disc ($r<<100$ au). Combining the lower CO abundance and the properties of the dust (e.g.~\citealt{vanLoon2005b, Oliveira2013, Jones2017b}) may reduce the likelihood of detecting CO bandhead emission. Less abundant CO produces fainter emission that is more easily swamped by the continuum. 
Less dust provides less shielding of CO which may further lower the abundance. \citet{Yasui2010} suggest that lower metallicity discs have a higher ionization fraction which may lead to higher accretion rates, especially if accretion is driven by the magnetorotational instability \citep[although this is not necessarily the case for higher-mass stars, see, e.g.][]{Kuiper2011}. 

Tantalisingly, a similar effect is seen in prominent ice bands in the spectra of massive YSOs in the SMC. \cite{Oliveira2011, Oliveira2013} found only an upper limit for CO ice.
This is most likely caused by a combination of lower gas-phase CO abundance, higher dust temperatures in the SMC YSOs, and the high interstellar radiation field or the cosmic-ray ionization rate affecting gas and grain-surface chemistry \citep[see e.g.,][]{Pauly2018}.  As both gas and ice CO signatures are weak this suggests lower CO abundance overall.

Variations in disc properties (e.g.~temperature, density etc.) with metallicity may also affect the CO bandhead emission.
For example, in hotter discs not only is the dust-sublimation region larger, but the higher temperatures and gas densities may lead to more of the CO rotational transitions near the bandhead becoming optically thick, further reducing the feature contrast \citep{Ilee2018b}. 
Observations of low-mass stars suggest shorter disc lifetimes in low metallicity regions \citep{Yasui2009,Yasui2010}. If this extends to higher-mass stars as well, stars in the SMC may spend a smaller fraction of their formation time in the narrow range of accretion rates where \citet{Ilee2018b} find the CO bandhead is most likely to be detected. 

Finally, the spectral resolution of the observations may also influence the detection rate of the first CO overtone, with rates lower for Galactic YSOs observed at high and low resolutions (R=35000-70000 and R=500) versus medium resolution (R=7000) data \citep[e.g][]{Hsieh2021}. Here the relatively weak CO first overtone emission maybe masked by the lower SNR or poor contrast above a dusty continuum. 
The spectral resolving power of the KMOS data is R$\sim$4000, with other sources in the Magellanic Clouds observed at resolutions between R=1000-17000. 
It is conceivable that, given the distances to the LMC and SMC the CO bandhead remains undetected especially if these sources have a relatively large continuum excess emission. A larger sample with better signal-to-noise ratio and greater spectral resolution is required to conclusively address this issue. 

The 1.083 $\mu$m feature is one of the brightest He\,{\sc i} lines in the IR.
As found by \citet{Reiter2019} and \citet{Ward2016, Ward2017} we also detect He~{\sc i} emission at higher rates ($>$40\%) than seen in high-mass Galactic YSOs; in these local samples the He\,{\sc i}~1.083 $\mu$m line is detected at a rate of 24\% \citep{Pomohaci2017}. Unfortunately,  \citet{Ward2016, Ward2017} did not observe their YSO samples in the Clouds at shorter near-IR wavelengths, and their only detections of He\,{\sc i} emission were of the 2.0587 $\mu$m line
(2$^1$P$^0$--2$^S $S) which is outside our spectral range. 
Nevertheless, our Magellanic Cloud results comparing the He~{\sc i} detection rates appear to be consistently higher than the observed rates for comparable He~{\sc i}~lines in the Milky Way \citep{Pomohaci2017, Cooper2013}. 
 Due to the small sample sizes these trends are tentative, and larger samples of massive-YSOs in both the Magellanic Clouds and the Milky Way are required to confirm any strong correlations with metallicity.
We speculate that these differences may arise due to the harder radiation YSOs experienced in the Magellanic Clouds, but the limited number of sources observed at both wavelengths prevents robust statistical analysis of the physical processes that may be responsible for these trends.

\section{Conclusions}
\label{sec:conclusion}

We have presented a detailed near-IR spectroscopic study with VLT/KMOS of the NGC 346 star-forming region in the SMC. We targeted {\em massive} YSO candidates that were photometrically identified by \citet{Sewilo2013} and \citet{Simon2007} to be in the early stages of formation  (Stages I -- III; \citealt{Robitaille2006}). 
We observed 18 YSO candidates in this region with KMOS, half of which were resolved into multiple objects in the continuum images from the IFUs.  From these data we extracted, medium-resolution YJ, H and K band spectra (with S/N $>$10) for twelve high-mass YSO candidates. 

The spectra are rich in features, particularly hydrogen recombination emission lines from the Paschen, Balmer and Pfund series. Nine of the 11 targets with K-band spectroscopy display Br$\gamma$ emission; H$_2$ and  He~{\sc i} are also detected at reasonably high rates ($\sim$45\%), confirming that these are massive stars in the early phases of formation.
No CO bandhead emission is seen in the K-band spectra, which is consistent with previous observations toward young stars in the Small Magellanic Cloud where weak CO bandhead emission has only been detected in one source \citep{Ward2017}. For comparison, the detection rate of the CO first overtone emission is 34\,\% in the Milky Way  when observed with similar spectral resolution \citep{Pomohaci2017}. 
Conversely, the detection of He\,{\sc i}~1.083 $\mu$m emission is detected at appreciably higher rates in our SMC sample than in the Galaxy. 
 Adding these new detections to other samples in the literature, provides the largest sample of medium-resolution YSO spectroscopy in the Magellanic Clouds,  totalling 53 sources, to date.  Thus providing better statistical evidence to explore the conditions experienced during the formation of high-mass stars in metal-poor environments compared to the solar neighbourhood.

We have also identified Y533A as an early O-type star and Y548A as slightly cooler star with an early spectral type, from their spectral absorption lines. Radial velocities of the targets are consistent with results for the young population of the SMC. Poor signal-to-noise prevented a detailed exploration of the fainter stars in the KMOS data cubes, with the impact of telluric features also limiting our analysis.  

NGC 346 is a target for a Cycle~1 program \citep{Meixner2017_jwst346} with the {\em James Webb Space Telescope} ({\em JWST}). With the greater spatial resolution and sensitivity of {\em JWST} we will push the study of active star formation in the cluster to solar-mass stars and test our hypothesis that metallicity affects the inner dust sublimation regions of dense accretion discs of massive YSOs, which is traced by the CO first-overtone emission around 2.3$\mu$m.

\section*{Acknowledgements}

We would like to thank the referee who's comments and suggestions significantly improved this paper. 
OCJ and MR received funding from the EU's Horizon 2020 program under the Marie Sklodowska-Curie grant agreement No 665593 awarded to the STFC. MR partially supported by an ESO fellowship. This research made use of Astropy,\footnote{http://www.astropy.org} a community-developed core Python package for Astronomy \citep{Astropy2013}. APLpy, an open-source plotting package for Python \citep{aplpy2012, aplpy2019}. 

\

\noindent {\it Facilities:}  VLT.

\section*{Data availability}

The data underlying this article will be shared on reasonable request to the corresponding author.


\input{journaldefs}


\input{ms_arxiv.bbl}



\appendix

\section{Continuum Images}
\label{sec:continiumnImages}

The continuum maps were generated using  the {\sc SpectralCube} \citep{spectralcube2016} python package and displayed with {\sc aplpy} using  a linear stretch, autoscaled to the peak flux.
In the 2MASS 6x catalogue \citep{Cutri2004} seven of the 18 original YSO candidates in our sample have matches to multiple sources within 1'', thus the detection of multiples in the KMOS data cubes is expected.
The RA and Dec.~positions of each source resolved and automatically detected using {\sc DAOStarFinder} in {\sc Photutils} ($>$5$\sigma$ detection above the background in the continuum flux maps) are given in Table~\ref{tab:specPositions},  sources $>$3$\sigma$ are listed in Table~\ref{tab:sourceALLPositions}. In instances where multiple sources were present, the brightest source in the cube was identified with an `A' suffix, with fainter sources designated `B'  and `C'. No cube had more than two sources with fluxes $>$5$\sigma$ above the background. Potential sources with $<$5$\sigma$ detections are not considered further.

\begin{table}
\centering
\caption{ List of all sources with a $>$3$\sigma$ detection as determined by {\sc DAOStarFinder} present in our KMOS datacubes. The wavelengths at which the source was detected is denoted by a check mark.}
\label{tab:sourceALLPositions}
\begin{tabular}{@{}lccccc@{}}
\hline
\hline
ID	&	RA	    &	Dec	   &	\multicolumn{3}{c}{S/N $>$3 Detection}\\
	& [J2000]	&	[J2000]	&	YJ	&	H &	K	\\
\hline
Y519A	&	00h58m40.240s	&	$-$72d11m22.19s	&	\checkmark	&	\ldots	&	\ldots	\\
Y523A	&	00h58m49.401s	&	$-$72d12m55.12s	&	\checkmark	&	\ldots	&	\ldots	\\
Y524A	&	00h58m53.567s	&	$-$72d12m54.78s	&	\checkmark	&	\checkmark	&	\ldots	\\
Y525A	&	00h58m55.641s	&	$-$72d08m31.99s	&	\checkmark	&	\checkmark	&	\checkmark	\\
Y525B	&	00h58m55.357s	&	$-$72d08m31.67s	&	\checkmark	&	\checkmark	&	\ldots	\\
Y525C	&	00h58m55.902s	&	$-$72d08m31.23s	&	\ldots	&	\checkmark	&	\ldots	\\
Y528A	&	00h58m57.148s	&	$-$72d12m35.36s	&	\checkmark	&	\checkmark	&	\checkmark	\\
Y528B	&	00h58m56.809s	&	$-$72d12m34.85s	&	\checkmark	&	\checkmark	&	\ldots	\\
Y532A	&	00h59m02.536s	&	$-$72d10m07.60s	&	\checkmark	&	\checkmark	&	\checkmark	\\
Y533A	&	00h59m02.923s	&	$-$72d10m34.90s	&	\checkmark	&	\checkmark	&	\checkmark	\\
Y535A	&	00h59m05.304s	&	$-$72d10m35.50s	&	\checkmark	&	\checkmark	&	\checkmark	\\
Y535B	&	00h59m05.664s	&	$-$72d10m35.04s	&	\checkmark	&	\ldots	&	\ldots	\\
Y538A	&	00h59m08.550s	&	$-$72d11m03.13s	&	\checkmark	&	\checkmark	&	\checkmark	\\
Y538B	&	00h59m08.833s	&	$-$72d11m03.66s	&	\checkmark	&	\checkmark	&	\ldots	\\
Y543A	&	00h59m11.654s	&	$-$72d12m23.74s	&	\checkmark	&	\checkmark	&	\checkmark	\\
Y544A	&	00h59m12.250s	&	$-$72d09m58.56s	&	\checkmark	&	\checkmark	&	\checkmark	\\
Y544B	&	00h59m12.333s	&	$-$72d10m00.25s	&	\checkmark	&	\checkmark	&	\checkmark	\\
Y545A	&	00h59m14.122s	&	$-$72d09m28.47s	&	\checkmark	&	\checkmark	&	\checkmark	\\
Y545B	&	00h59m14.230s	&	$-$72d09m26.88s	&	\checkmark	&	\checkmark	&	\ldots	\\
Y545C	&	00h59m13.828s	&	$-$72d09m27.70s	&	\ldots	&	\checkmark	&	\ldots	\\
Y547A	&	00h59m14.988s	&	$-$72d11m02.31s	&	\checkmark	&	\checkmark	&	\checkmark	\\
Y548A	&	00h59m17.046s	&	$-$72d09m14.86s	&	\checkmark	&	\checkmark	&	\checkmark	\\
Y549A	&	00h59m17.141s	&	$-$72d11m23.88s	&	\checkmark	&	\checkmark	&	\checkmark	\\
Y551A	&	00h59m19.209s	&	$-$72d09m30.29s	&	\checkmark	&	\checkmark	&	\checkmark	\\
Y551B	&	00h59m18.951s	&	$-$72d09m30.29s	&	\checkmark	&	\checkmark	&	\ldots	\\
Y553A	&	00h59m20.660s	&	$-$72d11m20.83s	&	\checkmark	&	\checkmark	&	\checkmark	\\
Y553B	&	00h59m20.421s	&	$-$72d11m22.19s	&	\checkmark	&	\ldots	&	\ldots	\\
Y556A	&	00h59m23.754s	&	$-$72d12m56.32s	&	\checkmark	&	\checkmark	&	\checkmark	\\
Y556B	&	00h59m23.953s	&	$-$72d12m55.12s	&	\checkmark	&	\checkmark	&	\ldots	\\
Y556C	&	00h59m23.702s	&	$-$72d12m54.78s	&	\checkmark	&	\checkmark	&	\ldots	\\
\hline
\end{tabular}
\end{table}

\section{Emission Line Fluxes}
\label{sec:em_lineFluxesApp}

Table~\ref{tab:linefluxes} provides summary of all detected spectral features for each object our NGC 346 KMOS sample, together with their measured emission-line fluxes. The measurement of the line fluxes is described in Section~\ref{sec:EmissionLineStrengths}. 
A question mark (?) indicates that the line identification is uncertain. The fluxes have not been corrected for extinction.
We do not provide line measurements for Pfund series emission and other lines (tentatively) identified in this region of the  spectrum due to the increased noise from atmospheric transmission. Similarly, if the line is present but in absorption an "A" is listed instead of a flux.

\begin{landscape}
\begin{table}
\small
\setlength{\tabcolsep}{2pt}
\centering
\caption{Common emission line fluxes towards the resolved-point sources in our NGC 346 KMOS sample, subdivided into the JY, H and K wavelength bands by the horizontal lines. Fluxes are given in W m$^{−2}$. 
"P" denotes that a line is present but a reliable estimate of its line flux is not possible due to the low contrast above (or poor fit to) the local continuum.  
"A" indicates that a line is present but is in absorption. 
The fluxes have not been corrected for extinction. 
Where a spectra was not obtained `--' is marked in the table. 
A `:' indicates that the line may have a potential P Cygni profile. }
\label{tab:linefluxes}
\begin{tabular}{@{}lcccccccccccc@{}}
\hline
\hline
Source ID	&	525A	&	532A	&	533A	&	535A	&	538A	&	544A	&	544B	&	545A	&	548A	&	549A	&	553A	&	556A	\\
Line	&	(Wm$^{-2}$)	&	(Wm$^{-2}$)	&	(Wm$^{-2}$)	&	(Wm$^{-2}$)	&	(Wm$^{-2}$)	&	(Wm$^{-2}$)	&	(Wm$^{-2}$)	&	(Wm$^{-2}$)	&	(Wm$^{-2}$)	&	(Wm$^{-2}$)	&	(Wm$^{-2}$)	&	(Wm$^{-2}$)	\\
\hline
He~{\sc i}	&	$<$1.7$\times 10^{-19}$	    &	2.08$\pm$0.12 $\times 10^{-19}$	&	A	                     &	:1.46$\pm$0.07$\times 10^{-18}$	&	--	&	1.51$\pm$0.07$\times 10^{-18}$	&	--	&	2.68$\pm$0.25$\times 10^{-20}$	&		    A               &	--	&	$<$6.3$\times 10^{-19}$	&$<$9.9$\times 10^{-20}$ 	\\
Paschen 6-3 &	$<$6.1$\times 10^{-20}$	    &	3.26$\pm$0.13 $\times 10^{-19}$	&	A	                     &	7.55$\pm$0.17$\times 10^{-18}$	&	--	&	5.33$\pm$0.13$\times 10^{-18}$	&	--	&	6.83$\pm$0.20$\times 10^{-19}$	&	$<$5.9$\times 10^{-21}$ &	--	&	5.78$\pm$0.21$\times 10^{-20}$	&	6.95$\pm$0.22$\times 10^{-19}$	\\
$[$Fe~{\sc ii}$]$& $<$ 2.3$\times 10^{-20}$	&	$<$1.9$\times 10^{-20}$ & $<$9.3$\times 10^{-20}$    &	7.25$\pm$0.64$\times 10^{-19}$	&	--	&	1.67$\pm$0.24$\times 10^{-19}$	&	--	&	5.56$\pm$1.30$\times 10^{-20}$	&	$<$3.1$\times 10^{-20}$	&	--	&	6.00$\pm$1.00$\times 10^{-20}$	&	$<$3.8$\times 10^{-20}$	\\
Paschen 5-3	&		    A?                  &	6.84$\pm$0.21 $\times 10^{-19}$	&	A	                     &	1.47$\pm$0.03$\times 10^{-17}$	&	--	&	8.46$\pm$0.23$\times 10^{-18}$	&	--	&	1.40$\pm$0.02$\times 10^{-18}$	&		    A               &	--	&	1.47$\pm$0.05$\times 10^{-18}$	&	1.28$\pm$0.04$\times 10^{-18}$	\\
\hline
Brackett 12-4	&	--	& $<$7.6$\times 10^{-18}$ P &	$<$1.1$\times 10^{-20}$	&	1.05$\pm$0.07$\times 10^{-18}$	&	--	&	4.67$\pm$0.14$\times 10^{-19}$	&	--	&	$<$8.5$\times 10^{-21}$	&	--	&	--	&	--	&	8.98$\pm$1.03$\times 10^{-20}$	\\
Brackett 11-4	&	--	& $<$7.1$\times 10^{-18}$ P &		 A                  &	1.27$\pm$0.10$\times 10^{-18}$	&	--	&	5.40$\pm$0.20$\times 10^{-19}$	&	--	&	$<$9.1$\times 10^{-21}$	&	--	&	--	&	--	&	9.95$\pm$0.55$\times 10^{-20}$	\\
H$_2$ 1-0 S(9)/Fe~{\sc ii}	&	--	&     A         &		 A                  &	2.99$\pm$0.45$\times 10^{-19}$	&	--	&	1.37$\pm$0.16$\times 10^{-19}$	&	--	&	1.79$\pm$0.08$\times 10^{-19}$	&	--	&	--	&	--	&	$<$3.5$\times 10^{-20}$	\\
Brackett 10-4	&	--	& $<$3.4$\times 10^{-18}$ P &		 A                  &	1.39$\pm$0.11$\times 10^{-18}$	&	--	&	6.94$\pm$0.23$\times 10^{-19}$	&	--	&	$<$7.5$\times 10^{-20}$	P &	--	&	--	&	--	&	1.16$\pm$0.07$\times 10^{-19}$	\\
\hline
H$_2$ 1-0:S(1)	& -- & 5.07$\pm$0.60 $\times 10^{-19}$   & A & 1.36$\pm$0.25$\times 10^{-18}$   & 1.39$\pm$0.24$\times 10^{-19}$	  & 4.13$\pm$0.76$\times 10^{-19}$   & 3.29$\pm$0.31$\times 10^{-19}$	  & $<$3.5$\times 10^{-20}$ & $<$2.8$\times 10^{-20}$ &	4.07$\pm$0.39$\times 10^{-19}$   &	$<$4.4 $\times 10^{-20}$ & 1.08$\pm$0.06$\times 10^{-19}$   \\
Brackett 7-4	& -- & 3.35$\pm$0.54 $\times 10^{-19}$   & $<$9.6$\times 10^{-21}$ & 4.46$\pm$0.16$\times 10^{-18}$   &$<$6.7$\times 10^{-20}$ P & 2.04$\pm$0.18$\times 10^{-18}$   & 1.82$\pm$0.18$\times 10^{-19}$	  &	2.50$\pm$0.11$\times 10^{-19}$   & $<$4.8$\times 10^{-20}$ &	5.20$\pm$0.40$\times 10^{-19}$	&	4.36$\pm$0.14$\times 10^{-19}$	 & 2.92$\pm$0.05$\times 10^{-19}$   \\
H$_2$ 1-0 S(0)	& -- & 1.47$\pm$0.61 $\times 10^{-19}$   & $<$3.1$\times 10^{-20}$ & $<$4.7$\times 10^{-19}$ & $<$4.7$\times 10^{-20}$  & 1.56$\pm$0.81$\times 10^{-19}$   & 5.78$\pm$1.20$\times 10^{-20}$	  & $<$1.7$\times 10^{-20}$ & 7.00$\pm$4.01$\times 10^{-20}$	  &	1.46$\pm$0.20$\times 10^{-19}$	&	$<$5.4 $\times 10^{-20}$ & $<$6.5$\times 10^{-20}$ \\
\hline
\end{tabular}
\end{table}
\end{landscape}

\bsp	
\label{lastpage}
\end{document}

%% file: journaldefs.tex

\def\aj{AJ}					
\def\actaa{Acta Astron.}                        
\def\araa{ARA\&A}				
\def\apj{ApJ}					
\def\apjl{ApJL}					
\def\apjs{ApJS}					
\def\ao{Appl.~Opt.}				
\def\apss{Ap\&SS}				
\def\aap{A\&A}					
\def\aapr{A\&A~Rev.}				
\def\aaps{A\&AS}				
\def\azh{AZh}					
\def\baas{BAAS}					
\def\jrasc{JRASC}				
\def\memras{MmRAS}				
\def\mnras{MNRAS}				
\def\pra{Phys.~Rev.~A}				
\def\prb{Phys.~Rev.~B}				
\def\prc{Phys.~Rev.~C}				
\def\prd{Phys.~Rev.~D}				
\def\pre{Phys.~Rev.~E}				
\def\prl{Phys.~Rev.~Lett.}			
\def\pasp{PASP}					
\def\pasj{PASJ}					
\def\qjras{QJRAS}				
\def\skytel{S\&T}				
\def\solphys{Sol.~Phys.}			
\def\sovast{Soviet~Ast.}			
\def\ssr{Space~Sci.~Rev.}			
\def\zap{ZAp}					
\def\nat{Nature}				
\def\iaucirc{IAU~Circ.}				
\def\aplett{Astrophys.~Lett.}			
\def\apspr{Astrophys.~Space~Phys.~Res.}		
\def\bain{Bull.~Astron.~Inst.~Netherlands}	
\def\fcp{Fund.~Cosmic~Phys.}			
\def\gca{Geochim.~Cosmochim.~Acta}		
\def\grl{Geophys.~Res.~Lett.}			
\def\jcp{J.~Chem.~Phys.}			
\def\jgr{J.~Geophys.~Res.}			
\def\jqsrt{J.~Quant.~Spec.~Radiat.~Transf.}	
\def\memsai{Mem.~Soc.~Astron.~Italiana}		
\def\nphysa{Nucl.~Phys.~A}			
\def\physrep{Phys.~Rep.}			
\def\physscr{Phys.~Scr}				
\def\planss{Planet.~Space~Sci.}			
\def\procspie{Proc.~SPIE}			
\let\astap=\aap
\let\apjlett=\apjl
\let\apjsupp=\apjs
\let\applopt=\ao
